\documentclass[10pt,journal]{IEEEtran}

\def\ps@headings{%
\def\@oddhead{\mbox{}\scriptsize\rightmark \hfil \thepage}%
\def\@evenhead{\scriptsize\thepage \hfil \leftmark\mbox{}}%
\def\@oddfoot{}%
\def\@evenfoot{}}
\makeatother
\usepackage{booktabs} 
\usepackage{url}
\usepackage{svg}
\usepackage{etoolbox}
\usepackage{cuted}
\usepackage{lipsum}
\usepackage{amssymb}
\usepackage{array}
\usepackage{amsmath}
\usepackage{tikz}
\usepackage{color}
\usepackage{dirtytalk}
\usepackage{algpseudocode}
\usepackage{graphicx}
\usepackage{caption}
\usepackage{mdframed}
\usepackage{tcolorbox}
\usepackage{comment}
\usepackage{mathtools}
\usepackage{pifont}
\usepackage{booktabs,makecell}
    
\usepackage{siunitx}

\usepackage[utf8x]{inputenc}
\usepackage{amsmath,amsthm}
\usepackage{multirow}
\usepackage[ruled,vlined]{algorithm2e}
\usepackage{fancyvrb}

\theoremstyle{definition}

\theoremstyle{plain}

\ifCLASSOPTIONcompsoc
  \usepackage[nocompress]{cite}
\else
  \usepackage{cite}
\fi
\ifCLASSINFOpdf
\else
\fi
\begin{document}
%
\title{A Hierarchical and Location-aware Consensus Protocol for IoT-Blockchain Applications}

\author{Hao Guo, ~\IEEEmembership{Member,~IEEE,}
        Wanxin Li*, ~\IEEEmembership{Member,~IEEE,}
        and Mark Nejad ~\IEEEmembership{Member,~IEEE}\\
      
\IEEEcompsocitemizethanks{
\IEEEcompsocthanksitem Manuscript received October 7, 2021; revised  January 19, 2022 and May 2, 2022; accepted May 17, 2022. \textit{(Corresponding author: Wanxin Li.)}  
\IEEEcompsocthanksitem Hao Guo is with the Research \& Development Institute of Northwestern Polytechnical University in Shenzhen, 518057, China (e-mail: haoguo@nwpu.edu.cn).
\IEEEcompsocthanksitem Wanxin Li is with the Department of Communications and Networking, Xi'an Jiaotong-Liverpool University, Suzhou, 215123, China (e-mail: wanxinli@udel.edu).
\IEEEcompsocthanksitem Mark Nejad is with the Department of Civil and Environmental Engineering, University of Delaware, Newark, Delaware, 19716, USA (e-mail: nejad@udel.edu).
}
}

%
%

\markboth{IEEE TRANSACTIONS ON NETWORK AND SERVICE MANAGEMENT 
,~Vol.~XX, No.~XX,~2022}%
{Shell \MakeLowercase{\textit{et al.}}: Bare Demo of IEEEtran.cls for Computer Society Journals}
%




\IEEEtitleabstractindextext{%
\begin{abstract}



Blockchain-based IoT systems can manage IoT devices and achieve a high level of data integrity, security, and provenance. However, incorporating existing consensus protocols in many IoT systems limits scalability and leads to high computational cost and consensus latency. In addition, location-centric characteristics of many IoT applications paired with limited storage and computing power of IoT devices bring about more limitations, primarily due to the location-agnostic designs in blockchains. We propose a hierarchical and location-aware consensus protocol (LH-Raft) for IoT-blockchain applications inspired by the original Raft protocol to address these limitations.
The proposed LH-Raft protocol forms local consensus candidate groups based on nodes' reputation and distance to elect the leaders in each sub-layer blockchain. It utilizes a threshold signature scheme to reach global consensus and the local and global log replication to maintain consistency for blockchain transactions.
 To evaluate the performance of LH-Raft, we first conduct an extensive numerical analysis based on the proposed reputation mechanism and the candidate group formation model. We then compare the performance of LH-Raft against the classical Raft protocol from both theoretical and experimental perspectives.
We evaluate the proposed threshold signature scheme using Hyperledger Ursa cryptography library to measure various consensus nodes' signing and verification time. Experimental results show that the proposed LH-Raft protocol is scalable for large IoT applications and significantly reduces the communication cost, consensus latency, and agreement time for consensus processing.

\end{abstract}
}
 \maketitle
 \pagestyle{headings} 
 \IEEEdisplaynontitleabstractindextext
 \IEEEpeerreviewmaketitle

\section{Introduction}
\label{sec:introduction}

Modern IoT networks are often large-scale, dynamically located, and globally distributed. By 2025, IoT devices such as smart home appliances, smartphones, and other types of smart sensors will increase to more than 75 billion\footnote{https://www.statista.com/statistics/471264/iot-number-of-connected-devices-worldwide/}. Many IoT networks require massive data communication and need to manage unreliable and failure messages, among others, automatically. Consensus protocols promise to achieve overall system reliability in the presence of inconsistent and failure messages by coordinating processes to reach agreements.  State machine replication (SMR) is a fundamental method for system availability and fault tolerance in distributed systems. For instance, Autopilot~\cite{isard2007autopilot} builds fault-tolerant replicas in Microsoft's data centers worldwide using consensus protocols. Google File System utilizes the Chubby~\cite{burrows2006chubby} lock service to reach the consensus for the replication of different files.  However, there are significant challenges in current IoT applications, including data integrity, resource-intensive consensus mechanisms, high latency, and limited scalability~\cite{putra2021trust,ZHU2019527}.


Blockchain is a distributed ledger that records transactions among multiple participants in a verifiable manner. Blockchain can reduce the costs involved in verifying transactions as a distributed ledger by removing the need for a trusted third-party operating as a centralized authority.  Since the introduction of Bitcoin~\cite{nakamoto2008bitcoin}, blockchain applications have expanded beyond cryptocurrencies and financial-related fields. The smart contract's\footnote{https://ethereum.org/} invention \cite{wood2014ethereum} leads to the development of more varied applications such as blockchain-based intelligent transportation systems (e.g., \cite{li2020blockchain, guo2020proof, li2021ICBC}) and smart health (e.g., \cite{guo2019access,guo2020icbc}). However, blockchains with a complex application layer and smart contracts can incur significant computation for transaction execution.
 
 {}
 
{Over the past few years, novel IoT-blockchain applications have attracted an increasing interest (e.g., Helium, Chronicled, and Atonomi)  due to the ever-growing advances in blockchains and their capabilities. Helium\footnote{https://whitepaper.helium.com/} uses blockchain to connect low-power IoT devices (such as microchips and routers) to the Internet. Chronicled\footnote{https://www.chronicled.com/lp/chargeback-errors-whitepaper/} combines blockchain and IoT products to deliver end-to-end supply chain management.
Atonomi~\cite{lao2020g} provides blockchain-inspired solutions such as immutable identity and reputation tracking for IoT applications. However, there are still significant challenges when deploying IoT applications along with blockchains due to the limited computing power and storage of existing IoT devices. In addition, many IoT applications require location-awareness in design, but the current blockchains have a location-agnostic design.}
Nodes from all regions are encouraged to participate in consensus and verification of transactions submitted from anywhere. 

\begin{table*}[t]
\centering
\caption{{Comparison between consensus protocol.} }
\label{tab: consensus table}
\begin{tabular}{cccccc}
Consensus & \thead{Type} & \thead{Throughput (TPS)} & \thead{Scalability} & \thead{Network Overhead} & Communication Complexity\\ \hline
PoW \cite{Jakobssonpow}   & Permissonless       & tens  & High & High & $O(N)$ \\ \hline
PoS \cite{vasin2014blackcoin}   & Permissonless       & hundreds  & High & High & $O(N)$ \\ \hline
PBFT \cite{castro1999practical}  & Permissoned       & thousands  & Low & High & $O(N^2)$ \\ \hline
Raft \cite{ongaro2014search}  & Permissoned       & thousands  & Low & Low & $O(N)$ \\ \hline
HotStuff \cite{yin2019hotstuff}  & Permissoned       & thousands  & Low & Low & $O(N)$ \\ \hline

\end{tabular}
\end{table*}

{Another critical limitation of most consensus protocols is the performance issue, which indicates the ability to increase the number of nodes in the network. The consensus algorithms can also impact blockchains' performance and transaction latency, throughput, and security. For a recent survey on consensus protocols for the blockchain networks, we refer the reader to~\cite{xiao2020survey}. Consensus protocols, such as Proof-of-Work (PoW)~\cite{nakamoto2008bitcoin} and Proof-of-Stake (PoS)~\cite{wiki:pos}, can handle massive communication between various nodes to reach consensus. These consensus protocols can feasibly scale for large-scale IoT networks. However, their system has a low throughput (TPS) and high network overhead~\cite{xiao2020survey}.}

In a public blockchain, nodes are allowed to join or leave the network without authentication and permission~\cite{li2020scalable}. Consequently, proof-based algorithms such as PoW~\cite{nakamoto2008bitcoin} and Proof-of-Stake (PoS)~\cite{vasin2014blackcoin} are widely used in many public blockchain applications. {PoW can handle massive nodes in a blockchain network with the mining process. However, the mining process requires significant time and computation power.}  In addition, these consensus protocols have other limitations, such as low transaction throughput and high latency. For instance, Bitcoin can only process about 7 Transactions Per Second (TPS), and Ethereum can process about 15 TPS. The transaction confirmation latency is about 10 minutes in Bitcoin and 15 seconds in Ethereum~\cite{cao2019internet}.
{Consequently, consensus protocols such as PoW cannot meet the response time requirements of many IoT applications due to their computational complexity and limited computation power~\cite{guo2020proof}.}

Unlike public blockchains, permissioned blockchains have the flexibility to relax some security assumptions of permissionless blockchains and utilize lighter consensus protocols such as {\it Paxos}~\cite{lamport2001paxos}, PBFT~\cite{castro1999practical}, and {\it Raft}~\cite{ongaro2014search}, which leads to reduced processing time and computational costs. 
However, they are not designed for large-scale IoT networks; {PBFT needs multiple rounds of communications between a leader node and all nodes to reach a consensus. For example, PBFT can scale to 128 nodes in the Hyperledger blockchain system as evaluated in~\cite{2017BLOCKBENCH}. This all-to-one communication is resource-intensive and increases latency. In contrast, Raft is the consensus protocol which is designed to be easy to understand. It's equivalent to Paxos consensus protocol in fault-tolerance and performance. The difference is that Raft is decomposed into independent sub-problems and addresses all major pieces needed for practical systems.} 



{Table \ref{tab: consensus table} shows a comparison, with respect to different performance metrics, of five consensus mechanisms; the PoW \cite{Jakobssonpow}, PoS~\cite{vasin2014blackcoin}, PBFT \cite{castro1999practical}, Raft \cite{ongaro2014search}, and HotStuff~\cite{yin2019hotstuff}. We
compare different types of blockchain architectures, throughput (TPS), scalability, network overhead, and communication complexity. Note here that scalability refers to the number of nodes the consensus algorithm can process in the system and implies an upper bound on network size.  If the protocol can support over 100 participants in
the network, then we conclude the scalability is high; otherwise, it is low. We also classify
the network overhead as high or low. High latency is in the magnitude of minutes or seconds, and low is in milliseconds.}

{Both Raft and HotStuff protocols have high throughput and low network overhead. However, HotStuff is based on the BFT consensus protocol which is a partially synchronized network~\cite{DBLP:journals/iacr/ChanPS18a}, and the upper bound of a message latency in HotStuff is unknown~\cite{yao2021survey}.  Our proposed hierarchical and location-aware consensus protocol can address the existing location-aware and scalability problems. }




{Many IoT-blockchain applications, such as for managing the electric power grid \cite{lu2021edge,gai2019iotj}, can benefit from a hierarchical architecture to reduce the consensus process and data communication time. A hierarchical multi-layer blockchain network can communicate within its sub-layers and achieve the consensus in a more efficient way~\cite{li2020scalable}. In this paper, we propose a novel consensus protocol for blockchain-based IoT applications: Location-based Hierarchical Raft algorithm (LH-Raft), which is inspired by the original Raft protocol~\cite{ongaro2014search}. 
By incorporating the geographic information of the IoT devices,  LH-Raft boosts the blockchain performance and makes the system more dynamic and  immune to malicious attacks~\cite{douceur2002sybil}. }

\begin{figure*}[t]
\centering
\includegraphics[width=0.87\textwidth]{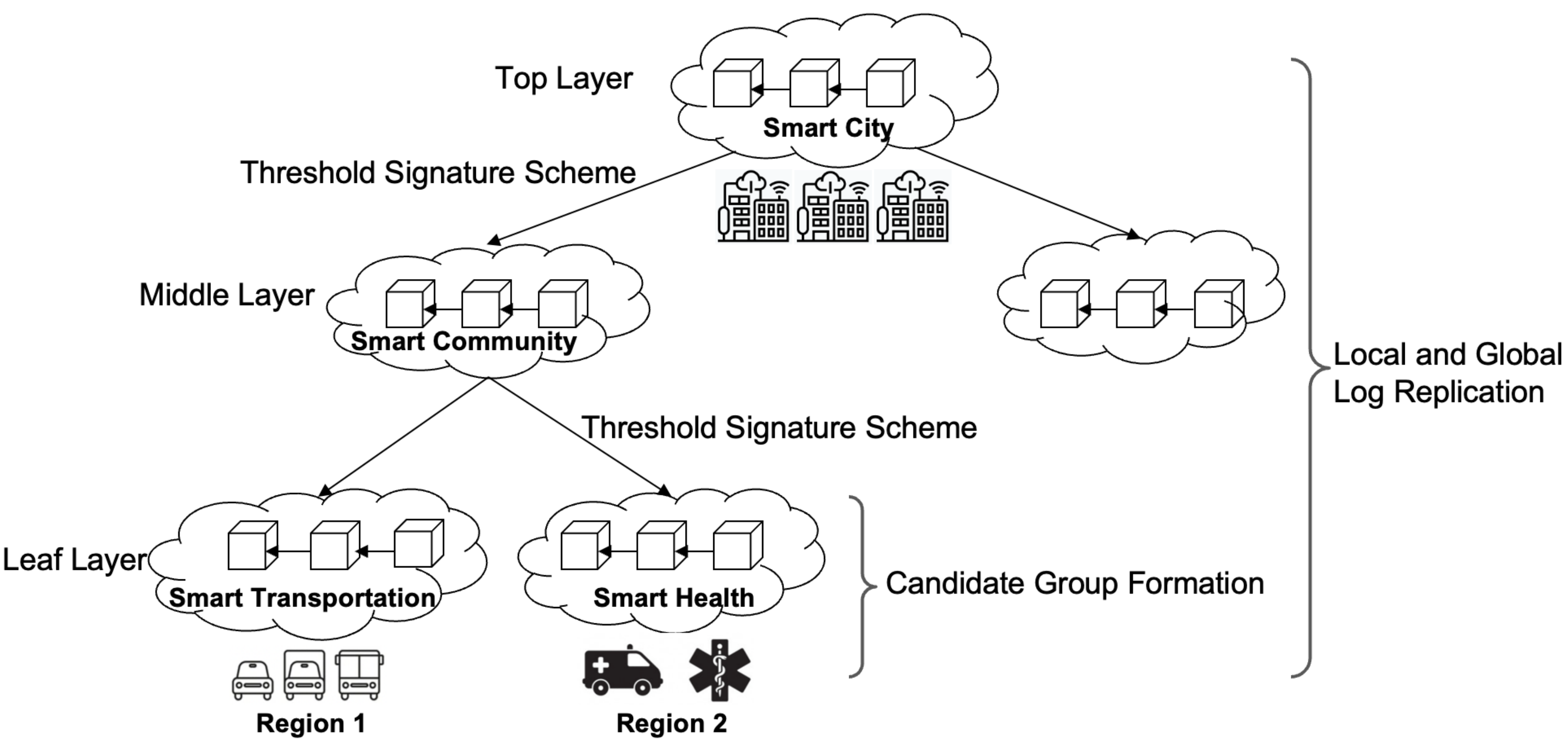}
 \caption { An overview of the system architecture.}
\label{fig:arch}
\end{figure*}

{As shown in Fig.~\ref{fig:arch},  LH-Raft engages a few {\it candidate nodes} with a low consensus latency in each sub-layer (e.g., can be leaf layer or middle layer) blockchain network, making the blockchain-based IoT system more efficient. As shown in the leaf layer, LH-Raft forms local candidate groups based on nodes' reputation and distance score to elect the {\it local leaders}. Next, all the {\it local leaders} from multiple-leaf layer blockchains and other {\it candidate nodes} will again elect the {\it upper leader} by utilizing the threshold signature scheme to reach the upper layer consensus. Note this process will happen once between the middle and top layers. In the end, it will reach the top layer and elects the {\it global leader}. Our proposed scheme partitions the blockchain network into a hierarchical structure based on the IoT device's regional information and utilizes the local and global log replication to maintain consistency for all blockchain transactions through multiple layers.}

%
%
%
%


This paper makes the following contributions: 
\begin{itemize}

\item  {We design LH-Raft consensus protocol, which constructs sub-layers  {\it local consensus} based on IoT devices' candidate group formation score, and builds a hierarchical structure by utilizing a threshold signature scheme with local and global log replication to reach {\it global consensus} and maintain consistency among 3-layer blockchain system.} 



 \item {We propose a reputation mechanism and candidate group formation model to engage consensus nodes.
 We design a threshold signature scheme, location-based hierarchical raft protocol, and local and global log replication scheme.  LH-Raft achieves higher transaction throughput with lower network overhead than the original Raft.} 

\item {We analyze LH-Raft 
 and compare it with the original Raft protocol. We conduct theoretical analysis regarding system performance, overhead, and fault-tolerance. 
 We simulate system performance with execution and communication cost, message passing, and consensus latency. We construct and evaluate the threshold signature scheme for the proof-of-concept model and execution time. }

\end{itemize}
The rest of the paper is organized as follows. We discuss the related work in Section II. Section III presents the background knowledge for the Paxos algorithm, Raft protocol, geographic information, and bilinear pairing-based cryptography. In Section IV, we describe the system architecture. Specifically, we present the candidate group formation model, threshold signature scheme, location-based hierarchical raft protocol (LH-Raft), and the local and global log replication scheme. Section V analyzes the LH-Raft protocol from three perspectives: performance, overhead, and fault-tolerance. In Section VI, we describe the theoretical properties and our prototype of LH-Raft and conduct experiments to evaluate the proposed scheme with network consistency. In Section VII, we conclude the paper and point out promising future research directions.

\section{Related Work}



In this section, we provide the related work for hierarchical consensus protocols and IoT-blockchain applications.

Yu et al.~\cite{yu2020layerchain} proposed a hierarchical edge-cloud blockchain architecture named LayerChain. They described a layered structure to save the blockchain transaction data in multiple distributed clouds and edge nodes. 
{Chuang et al.~\cite{chuang2020hierarchical} proposed a hierarchical blockchain-based data service platform in MEC environments. This system provided an adaptive PoW consensus scheme that dynamically changed the hash puzzle's difficulty and enhanced resource-constrained IoT devices. Zhang et al.~\cite{zhang2020blockchain} proposed a blockchain-based trust management system for IoV, which utilizes the consensus mechanism that integrates PoW and PoS to ensure all vehicles with a large change in reputation could be updated in the blockchain. Cui et al.~\cite{cui2021secure} proposed one secure and efficient data sharing mechanism among vehicles based on a consortium blockchain. They described an enhanced delegated proof-of-stake (DPoS) consensus protocol based on the trust score model.}

Yang et al.~\cite{yang2019blockchain} proposed a hierarchical trust networking architecture to implement JointCloud (HTJC). By developing the credit bonus-penalty strategy (CBPS), HTJC can address the trust issue and provide participants with a secure and trusted trade environment. 
Hou et al.~\cite{hou2021intelligent} described  an intelligent transaction migration scheme for the RAFT-based blockchain in IoT applications to migrate transactions in busy areas to idle regions intelligently and reduce the network latency significantly. Fu et al.~\cite{fu2021improved} proposed the AdRaft, which optimizes the original Raft consensus protocol for the Hyperledger Fabric platform in terms of both log replication and leader election phases. Xu et al.~\cite{xu2020blockchain} proposed the blockchain-based data auditing scheme and designed a client-side data deduplication scheme based on bilinear-pair techniques to reduce the burden on service providers and users. 

Lao et al.~\cite{lao2020g} proposed G-PBFT (Geographic-PBFT), a location-based and scalable consensus mechanism for IoT-blockchain applications. In their design, G-PBFT utilized the era switch mechanism to maintain the dynamics in the IoT devices. The experiment results showed that G-PBFT reduced the network overhead and consensus time significantly. 
Li et al.~\cite{li2020scalable} proposed a scalable multi-layer PBFT consensus protocol for blockchain. The proposed double-layer PBFT scheme reduces the communication complexity significantly. They also analyzed the security threshold based on the faulty probability determined and the faulty number determined models. An et al.~\cite{an2019tcns} proposed a decentralized privacy-preserving model based on the twice verifications process and consensuses of the blockchain system. They introduced a twice consensus mechanism, ensuring that data can be traced and prevented from being impersonated and denied.

{Kantesariya proposed a sharding scheme and validation protocols for a hierarchical blockchain architecture called OptiShard. In this scheme,  network nodes are divided into multiple disjoint shards, and the majority of transactions are distributed among these shards in a non-overlapped way~\cite{Kanicbc}. To address the storage issue, Wang et al.~\cite{wangblockchain} proposed an architecture that features a hierarchical storage structure where the majority of the blockchain is placed in the clouds, and the most recent blockchain transactions are stored in the overlay network of the individual IoT networks. Lu et al.~\cite{lu2021edge} proposed the edge-blockchain lightweight privacy-preserving data aggregation for smart grid application, which integrated edge computing and blockchain to formulate a three-layer architecture. Chai et al.~\cite{chai2020hierarchical} proposed a hierarchical blockchain-enabled federated learning algorithm for knowledge sharing in Internet of Vehicles, which builds a light Proof-of-Knowledge (PoK) consensus mechanism. However, the blockchain system performance measurement is missing.}


Lin et al.~\cite{lin2020blockchain} presented a Peer-to-Peer (P2P) computing resource trading scheme to balance the computing resource spatio-temporal dynamic demands in IoV-assisted smart city, and constructed a consortium blockchain approach and demonstrated the process of secure computing resource trading without involving a centralized trusted third-party. Liu et al.~\cite{liu2020sshc} proposed a secure and scalable hybrid consensus protocol for sharding blockchains with the formal security framework, and they designed a pipelined Byzantine fault tolerance scheme for the intra-shard consensus. Berger et al.~\cite{berger2020aware} introduced a novel mechanism that can improve the geographical scalability of consensus with nodes being widely spread across the real world. Their protocol is an automated and dynamic voting weight tuning and leader positioning scheme, which supports the emergence of fast quorums in the system.

This research work is the first effort to propose a blockchain-inspired hierarchical architecture with location-aware consensus in IoT-blockchain applications to the best of our knowledge. We also proposed a new location-based hierarchical raft protocol, compared the system performance with the classical raft protocol, and experimented with the threshold signature scheme.

\section{Background Knowledge}
This section briefly introduces the Paxos protocol, Raft protocol, geographic information's basics, and bilinear pairing-based cryptography. 

\subsection{Paxos Algorithm}

The Paxos algorithm was first introduced by Lamport in 1989 and later explained in the paper {\it Paxos made simple, 2001}~\cite{lamport2001paxos}. Paxos is an algorithm used to achieve consensus among distributed nodes that communicate through an asynchronous network. One or more nodes propose a value to Paxos, and the consensus is reached when the majority of nodes running Paxos agree on one of the proposed values. 

The Paxos algorithm executes as follows:
The {\it proposer} sends a message prepare({\it n}) to all {\it accepters}. 
Every accepter will compare {\it n} with the highest-numbered proposal for which it has responded to the prepared message. If {\it n} is greater, it will react with the $ack(n, v, n_v)$ where v is the highest-numbered proposal that has been accepted, and $n_v$ is the number of the proposal. 
The {\it proposer} will wait to receive $ack$ from the majority of {\it accepters}. If any $ack$ contained a value, it would set $v$ to the most recent (number ordering within the proposal) value received. Next, it will send $accept(n, v)$ message to all $accepters$.
Upon receiving $accept(n, v)$, an {\it accepter} accepts $v$ unless it has received $prepare(n')$ for some existing $n' > n$. If the majority of acceptors accept the value, then this value becomes Paxos protocol's decision value~\cite{lamport2001paxos}.

{Besides the Basic Paxos, Multi-Paxos can have the graphic representation of the flow messages, while the Cheap-Paxos extends the Basic Paxos to tolerate {\it f} failures with {\it f+1} main processors and {\it f} auxiliary processors with dynamic reconfiguration after each failure~\cite{lamport2004cheap}. Fast Paxos generalizes Basic Paxos to reduce the end-to-end delay from 3 to 2 messages in the client request. Byzantine Paxos adds an extra verification message which can act to distribute knowledge and verify the actions of other nodes.}

{In real-world scenarios, Google utilizes the Paxos protocol in the Chubby distributed lock service to keep the replicas consistent~\cite{castiglia2020hierarchical}. Microsoft uses the Paxos in their Autopilot cluster management service from the Bing application~\cite{castiglia2020hierarchical}.} Neo4j's graph database recently implemented the Paxos to replace the previous Apache ZooKeeper, and the Amazon Elastic Container Service utilizes Paxos to keep a consistent view of different cluster states. However, the Paxos protocol is complicated to implement and does not scale well in the large distributed network regarding the communication cost and latency time.



\subsection{Raft Protocol}
Raft is a consensus algorithm proposed in 2014 by Diego Ongaro~\cite{ongaro2014search}. From the fault-tolerance and system performance perspectives,  it is equivalent to Paxos. The Raft protocol decomposed the Paxos algorithm into independent sub-problems and explained sub-problems concisely and explicitly. Raft protocol states that every node in the replicated state machine can participate in any three states: {\it follower}, {\it candidate}, and {\it leader}.  One node can participate in any one of the above three states. Only the {\it leader} node can interact with the client; any request sent to the {\it follower} node is redirected to the {\it leader} node. However, a {\it candidate} can request votes to become the {\it leader}, and a {\it follower} node will only respond to the {\it candidates} or the {\it leader}~\cite{ongaro2014search}.
The Raft protocol divides time into short terms of arbitrary length. Each term is identified by a monotonically increasing number, which is named the term number.

{The consensus problem in classical Raft is decomposed into two independent sub-problems: {\it Leader Election} and {\it Log Replication}. When the current leader fails or the protocol initializes, a new leader needs to be elected. The {\it leader node} will send a heartbeat message to express domination to other {\it follower node}. The {\it leader node} is also responsible for the {\it log replication}. It will accept client requests. Each client request consists of a command to be executed by the replicated state machines in the cluster. Raft algorithm utilizes two types of Remote Procedure Calls (RPCs) to perform actions:
{\it RequestVotes} is sent by the {\it candidate nodes} to gather votes within the election procedure, and the {\it AppendEntries} is utilized by the {\it leader node} to replicate the log entries and also is served as the heartbeat message to determine if the node is still alive or not\footnote{http://thesecretlivesofdata.com/raft/}. The heartbeat message does not contain any log entries.}


{Raft protocol can guarantee the following safety properties: Election safety, which indicates that at most one {\it leader} can be elected in a given term. Leader append-only, a leader can only append the new entries to the existing logs (cannot overwrite or delete the entries).  For the state machine safety, if the existing server has applied the log entry to its state machine state, then no other server may apply a different command for the same log~\cite{ongaro2014consensus}, and the state machine safety is guaranteed by the restriction on the leader election procedure.}


\subsection{Geographic Information}
IoT devices' geographic information and timestamps can form local consensus nodes and prevent multiple types of attacks.  In our proposed consensus protocol, the geographic information includes coordinates (i.e., longitude and latitude) and timestamps collected by, for example, cell towers or navigation systems (e.g., GPS). To this end, we use  {\it \{longitude, latitude, timestamp\}} as the format for geographic information.

Coordinates, along with the timestamps, are widely used in many real-world applications. For instance, by tracking the smartphone's GPS information, a location-based service (LBS) like Lyft can provide ridesharing services. In addition, multiple recommendation services can also be offered based on location information, such as finding/suggesting nearby restaurants and shopping malls. Another example is the location-based parking lot services; they can show the real-time available parking spots and reduce cruising for parking; hence, reducing traffic. 

\subsection{Bilinear Pairing-based Cryptography}

$\mathbb{G}_1$ and $\mathbb{G}_2$ are two multiplicative cyclic groups. For both cyclic groups, the prime order is $p$, and the generator of $\mathbb{G}_1$ is $g$. There exists a mapping function $e$: $\mathbb{G}_1\times \mathbb{G}_1 \rightarrow \mathbb{G}_2$, with an efficient algorithm, which, for all $g_1, g_2 \in \mathbb{G}_1$, we can compute $e(g_1, g_2)$. Map $e$ is termed bilinear if it has the following two properties:


1) \textbf{Bilinearity:} $\forall$ $g_1, g_2 \in \mathbb{G}_1$ $\And$ $a, b \in \mathbb{Z}_p$, $\exists$ $e(g_1^a, g_2^b) = e(g_1, g_2)^{ab}$.

2) \textbf{Non-degeneracy:} $e(g, g) \neq 1$.

Pairing-based cryptography utilizes a pairing function between elements of two cryptographic groups to a third group with the mapping operation: $e: \mathbb{G}_1 \times \mathbb{G}_2 \longrightarrow \mathbb{G}_T$ to establish the cryptographic system. 
Given $g^z$, we can examine if $g^z = g^{xy}$ without revealing any actual information of $x, y, z$, by checking that if $e(g^x, g^y) = e(g, g^z)$ holds. By utilizing the bilinear property $x+y+z$ times, we can determine that whether $e(g^x, g^y) = e(g,g)^{xy} = e(g,g)^z = e(g,g^z)$. In consequence, $\mathbb{G}_T$ is a prime order group, so that $xy = z$ will hold.

\section{System Architecture}
\begin{figure}[t]
\centering
\includegraphics[width=0.48\textwidth]{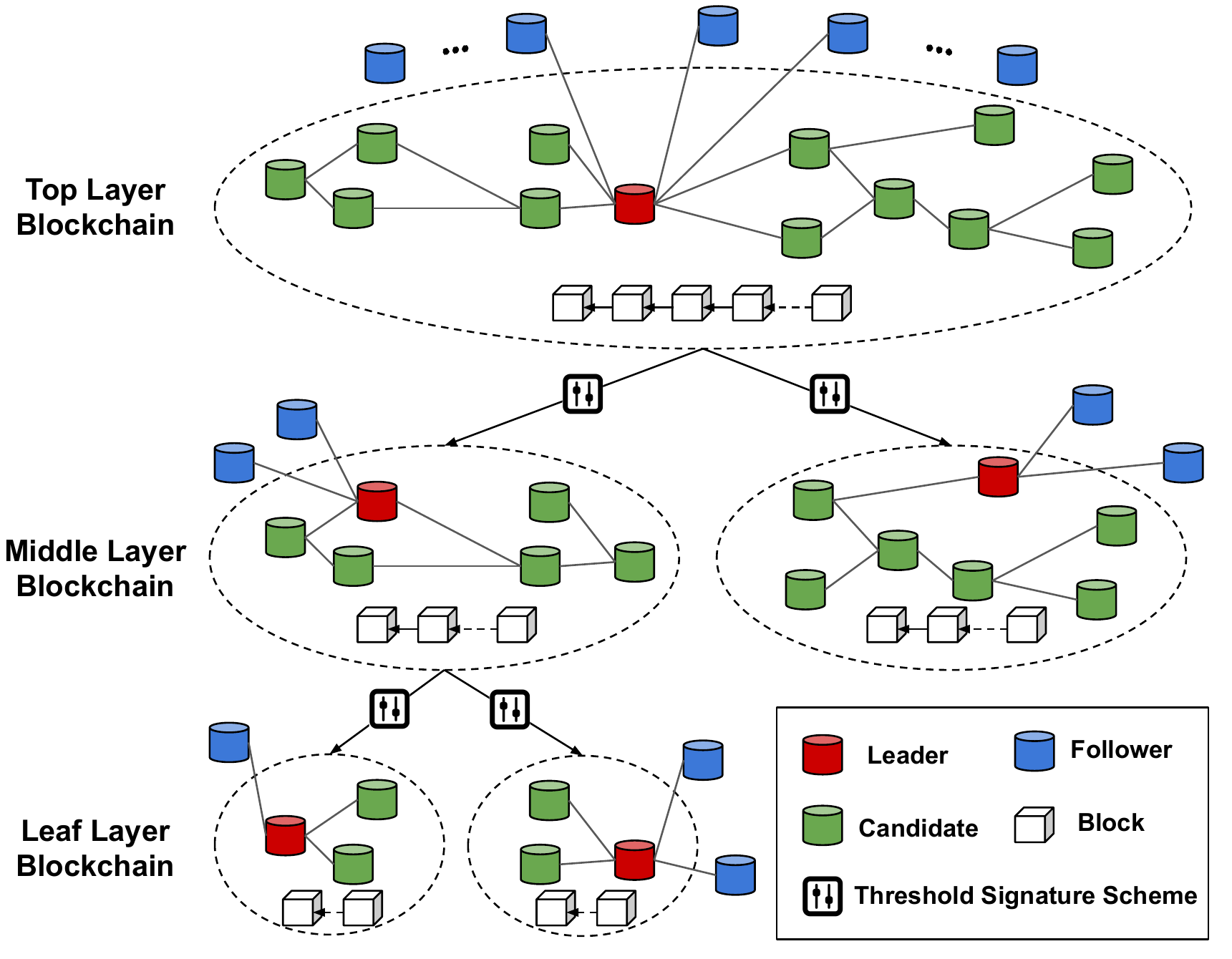}
\caption{Location-based hierarchical blockchain system for IoT applications.}
\label{fig:hierarchical}
\end{figure}

The underlying location-centric characteristics inherent in many IoT applications necessitates location-aware design solutions. This section describes our proposed system 
architecture for location-aware consensus in IoT networks.  
In our approach, 
the global network is divided into sub-blockchains based on regional information. These sub-blockchains are connected in a hierarchical structure forming a more extensive global blockchain network. The system provides a multi-chain and multi-level structure, as shown in Fig.~\ref{fig:hierarchical}. 
We first define the following entities that take part in the proposed architecture.

\begin{itemize}
\item Blockchain: Blockchain serves as the coordinator for IoT devices and manages data sharing and access activities. It forms a dynamic hierarchical structure based on the IoT device's geographic information, including top, middle, and leaf layers.

{\item IoT Nodes: The IoT nodes participate in the consensus process. There are three types of IoT nodes: {\it Follower}, {\it Candidate}, and {\it Leader} nodes. They can switch roles seamlessly between these statuses. All {\it Candidate} nodes together form the consensus nodes group, which elects the {\it Local Leader} node. Next, all {\it Local Leader} nodes and other {\it Candidate} nodes elect the {\it Global Leader} by utilizing the threshold signature scheme. Note that the {\it Leader} node in the top layer blockchain is the {\it Global Leader}.}

\item Client: A client node only requests new transactions to append data to the ledger, and they do not participate in the consensus procedure. For example, a healthcare system's smart devices can host client nodes to request electronic health record updates. 

\item Threshold Signature Scheme: The threshold signature scheme is proposed to achieve consensus among hierarchical blockchain layers in the architecture. We will explain the detailed construction in the following subsections.


\end{itemize}

\begin{figure}[t]
\centering
\includegraphics[width=0.487\textwidth]{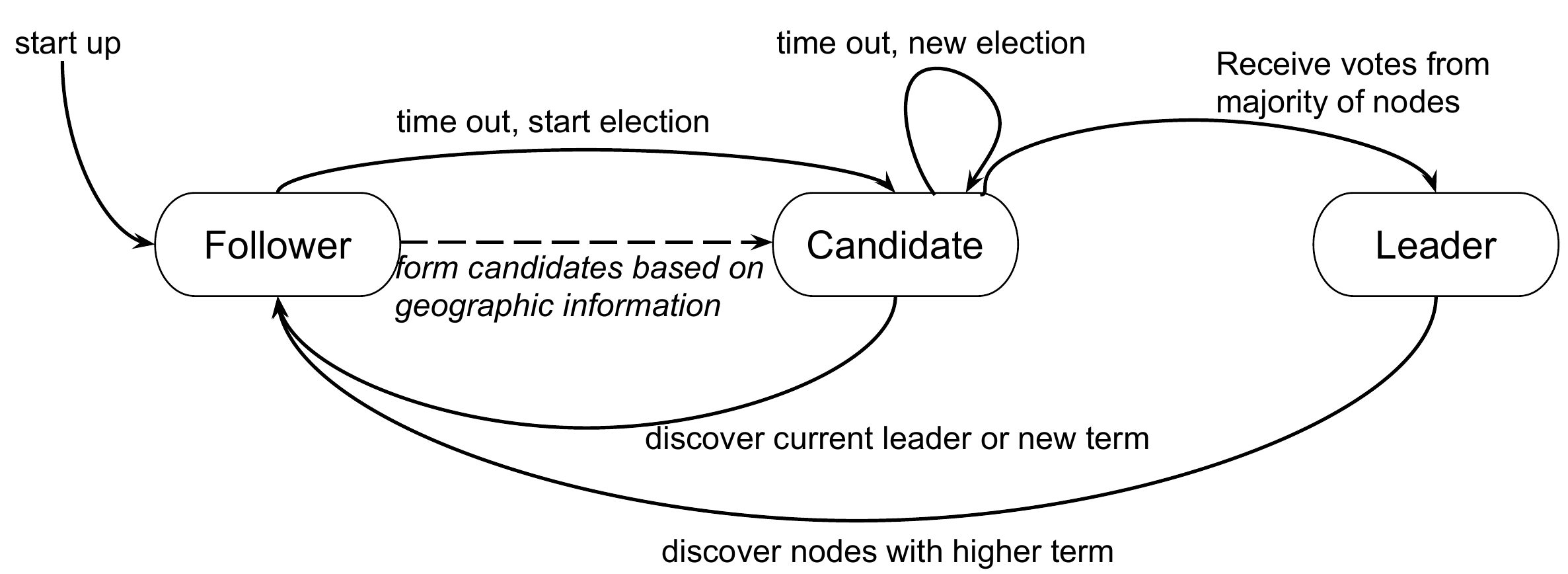}
\caption{Location-based hierarchical Raft protocol.}
\label{fig:lhraft}
\end{figure}

As shown in Fig.~\ref{fig:lhraft}, the proposed location-based hierarchical raft protocol has three participant entities: {\it Follower}, {\it Candidate}, and {\it Leader}. These interconvertible nodes can change their status to other roles. The {\it Candidate} and {\it Leader} nodes participate in the consensus process. The {\it Leader} nodes maintain the integrity and confidentiality of the blockchain system and broadcast the newly generated transactions to the {\it Follower} nodes. By contrast, the {\it Follower} nodes will only start new election processes and form {\it Candidate} nodes groups based on their geographic information. Transactions are determined among {\it Candidate} and {\it Leader} nodes to reduce communication overhead. If any message is failed, a node will resend the message again after the timeout period. The role of a node in our proposed scheme is not fixed, a {\it Follower} node can become the {\it Candidate} node, and {\it Candidate} node can become the {\it Leader} node. On the other hand, if the location of a {\it Leader} node has been changed or it conducts malicious action, it can be detected by the voting process by the {\it Candidate} nodes. 

The proposed architecture is designed in a way that is not affected by the size of the IoT network. Rather than all nodes participating in the consensus procedure, nodes execute {\it local consensus} within each sub-layer blockchain. Each sub-layer blockchain leader participates in the hierarchical consensus by utilizing the threshold signature scheme and local/global log replication scheme to reach {\it global consensus}.

{In the remainder of this section, we first describe a reputation mechanism incorporated in an optimal candidate group formation. Next, we present the threshold signature scheme to reach the consensus between multiple blockchain layers in the hierarchical architecture. We describe the location-aware hierarchical raft protocol with detailed constructions. Finally, we discuss the local and global log replication process for inter-layer and intra-layer network consistency.}


\subsection{Candidate Group Formation}

A location-based candidate group formation that dynamically forms a controlled number, $M$, of reliable local nodes for consensus processing is essential in lowering the computational complexity of consensus processing of location-centric IoT applications. {The candidate group formation process is invoked when there is a transaction request, and therefore the system needs to run a consensus process.}  
We propose to optimize candidate group formation regarding their reputation and distance, with the goal of achieving significantly increased throughput and reduced latency and consumed system resources (e.g., bandwidth) compared to global nodes. The reputation mechanism is an integral part of the system to engage reliable nodes and safeguard system integrity. In the reputation mechanism, nodes participating in the consensus processing leading to a new block generation increase their reputation.   
On the other hand, the reputation score of malicious nodes such as the ones causing a fork will be decreased. 

We introduce \emph{reputation graph}~$(\mathcal{V}, A)$ to model the reputation of nodes, 
where~$ \mathcal{V}$ and~$A$ are sets of nodes and arcs, respectively. 
The weight~$\omega(\mathcal{V}_i,\mathcal{V}_j)$~$\in~[0,R]$ associated with arc~$(\mathcal{V}_i, \mathcal{V}_j)$ represents the $\emph{reputation}$ that ~$\mathcal{V}_i$ assigns to~$\mathcal{V}_j$, 
 based on~$\mathcal{V}_i$'s record of~$\mathcal{V}_j$'s past performance. 
{ Reputation assignments to new nodes are one and $ \mathcal{R}$ is chosen by the system designer and corresponds to the highest reputation that a node can assign to another node.} Note that the \emph{reputation graph} does not have self-arcs (i.e., $\nexists$ $(\mathcal{V}_i, \mathcal{V}_i) \in A$), which means nodes cannot assign reputation score to themselves. 

We define the \emph{normalized reputation} that~$\mathcal{V}_i$ assigns to~$\mathcal{V}_j$, ~$\rho(\mathcal{V}_i, \mathcal{V}_j)\in~$[0,1], 
by dividing~$\omega(\mathcal{V}_i,\mathcal{V}_j)$ to sum of all reputation scores that ~$\mathcal{V}_i$ assigns as follows:

\begin{equation}
\rho\mathcal({V}_i, \mathcal{V}_j) = \frac{\omega(\mathcal{V}_i, \mathcal{V}_j)}{ \sum_{\mathcal{V}_k\in \mathcal{I}(\mathcal{V}_i)}{\omega(\mathcal{V}_i, \mathcal{V}_k)}}\label{eq:nt},
\end{equation}
where $\mathcal{I}(\mathcal{V}_i) = \{\mathcal{}V_j | \exists (\mathcal{V}_i, \mathcal{V}_j) \in A \}$ is the set of nodes that $\mathcal{V}_i$ has interacted with in the past.
Note that summation over all normalized reputation scores that~$\mathcal{V}_i$ assigns is one: 
\begin{equation}
\sum_{\mathcal{V}_j \in \mathcal{V}, \forall j\neq i}{\rho(\mathcal{V}_i, \mathcal{V}_j)} = 1.
\end{equation}

The reputation score of each node~$\mathcal{\tau}_i$ is calculated by summation over all normalized reputation scores that are given to node~$\mathcal{V}_i$ as follows:

\begin{equation}
\mathcal{\tau}_i=\sum_{\mathcal{V}_j \in \mathcal{V}, \forall j\neq i}{\rho(\mathcal{V}_j, \mathcal{V}_i)}.
\end{equation}


{The reputation scores of nodes that are involved in the process will be updated based on the new reputation scores assigned by the nodes involved in that iteration. Note that if a node is not involved in the iteration, its reputation score remains the same.}

We define the distance score of each node~$\mathcal{\sigma}_i$ as follows:
\begin{equation}
\mathcal{\sigma}_i = \frac{\sum_{\mathcal{V}_i \in \mathcal{V}}{d_{\mathcal{V}_i}}}{ \mathcal{V} \times d_{\mathcal{V}_i}},
\end{equation}
where $d_{\mathcal{V}_i}$ is the distance of node $\mathcal{V}_i$ from the application-specific event{, which is updated based on the location information at the time of invoking the process,} and $ \mathcal{V}$ is the size of set~$ \mathcal{V}$, the total number of nodes. For example, in the application of auto accident forensics, one would like to be able to utilize the blockchain to optimally identify a subset of nodes local to the accident site, which can verify and come to a consensus on the spatial details of the event such as trajectories and coordinate locations.

We formulate the candidate group formation (CGF) as an Integer Program (IP), called IP-CGF. We define the following decision variables:

\begin{equation}
\begin{split}
X_{\mathcal{V}_i} =
\begin{cases} 
1 & \text{if node $i$ is assigned to the candidate group,}\\ 
0 & \text{otherwise.} 
\end{cases}
\end{split}
\end{equation} 

We formulate IP-CGF as follows:

\begin{equation}
{\text{Maximize} \sum_{\forall i} (\mathcal{\alpha\tau}_{i} + \mathcal{\beta\sigma}_i)  X_{\mathcal{V}_i} ,   \label{eq:ip2obj}}
\end{equation}

\text{Subject to: } 

\begin{align}
\label{eq:ip2c0}
\quad&\sum_{\mathcal{V}_i \in \mathcal{V}} X_{\mathcal{V}_i}  = M,\\
\label{eq:ip2c1}
\quad& X_{\mathcal{V}_i} \in \{0, 1\}. 
\end{align} 

The objective function (\ref{eq:ip2obj}) is 
to maximize the combination of the candidate group's reputation score as well as geographical closeness. {The objective function nonnegative coefficients, $\mathcal\alpha$ and $\mathcal\beta$, allow system designers to give different weights to reputation and distance based on application-specific needs.}
Constraint~(\ref{eq:ip2c0}) ensures that IP-CGF selects a pre-specified number of nodes, $M$, to be included in the candidate group. $M$ represents the controllable number of nodes in the consensus processing, and it can be varied depending on specific IoT applications.  Constraints~(\ref{eq:ip2c1}) specify that the decision variables are binary. In Section V, we conduct numerical analysis to evaluate the impact of changes in the candidate group size and the total number of participating nodes on the execution time of the candidate group formation mechanism.

\subsection{Threshold Signature Scheme}
We proposed the threshold signature scheme for trust-based hierarchical coordination and operation between two blockchain layers. For example, assume that a lower layer blockchain network contains $n$ consensus nodes, and the threshold is set as $t$-out-of-$n$ between the lower layer and the upper layer blockchain network. The upper layer blockchain, acting as the verifier network, will trust this lower layer blockchain only if at least $t$ consensus nodes' signatures are verified as legitimate. In addition, utilizing bilinear pairing-based cryptography, the signatures can be verified without disclosing any sensitive information.

\begin{algorithm}[ht]
\label{alg: zkp-keygen}
\SetAlgoLined
\LinesNumbered
\SetKwInOut{Input}{Input}
\SetKwInOut{Output}{Output}
\Input{For each consensus node $i$}
\Output{Verifier key $v_i$}
The permission issuer selects a random $a_i \in \mathbb{Z}_p$\ for consensus node $i$ \;
The permission issuer computes the verifier key as $v_i = g^{a_i} \in G$ \;
The permission issuer returns $v_i$ \;
\caption{Key Generation}
\end{algorithm}

\begin{algorithm}[ht]
\label{alg: zkp-proofgen}
\SetAlgoLined
\LinesNumbered
\SetKwInOut{Input}{Input}
\SetKwInOut{Output}{Output}
\Input{Each consensus node's MAC address $m_i$}
\Output{One-time signature $\delta_i$}
The system computes a hash digest $h_i$ based on MAC address and location information $m_i$ via \cite{rachmawati2018comparative}, as $h_i = H(m_i)$ \;
The system generates the one-time signature $\delta_i = {h_i}^{a_i} \in G$ \;
The system sends $\delta_i$ to the upper layer network \;
\caption{Threshold Signature Generation}
\end{algorithm}

\begin{algorithm}[ht]
\label{alg: zkp-proofverify}
\SetAlgoLined
\LinesNumbered
\SetKwInOut{Input}{Input}
\SetKwInOut{Output}{Output}
\Input{One-time signature $\delta_i$, hashed MAC address $h_i$, verifier key $v_i$}
\Output{Identity verification result $r$}
k = 0;

\For{
each one-time signature $\delta_i$
}{\eIf{$e(\delta_i, g) == e(h_i, v_i)$}{$r_i = True$ \; $k = k + 1$\;}{$r_i = False$ \;}}
the system checks
\eIf{$k \geq t$}{$r = True$ \;}{$r = False$ \;}
The system returns $r$ \;
\caption{Threshold Signature Verification}
\end{algorithm}

\vspace{3mm}
\noindent

We describe the procedure of the proposed threshold signature scheme in Algorithms 1, 2, and 3. The algorithms contain three main functions: Algorithm 1 shows the key generation function by the system administrator; Algorithm 2 describes the threshold signature generation process by the lower layer blockchain nodes; Algorithm 3 presents the threshold signature verifying function by the upper layer blockchain nodes, and a threshold $(t,~n)$ is considered in verifying the lower layer blockchain network. The construction of our proposed threshold signature scheme is shown below:

 \textit{Initial Setup}:  The system has the bilinear pairing function $e$: $\mathbb{G}_1\times \mathbb{G}_2 \rightarrow \mathbb{G}_T$, the secure hash function $H:M \rightarrow \mathbb{G}_1$, and the $(\mathbb{G}_1, \mathbb{G}_2, \mathbb{G}_T, e, g_1, g_2, p, h)$ represents the public parameters.

\textit{Key Generation}: A trusted authority generates signing-verifying key pairs for all consensus nodes in this step. The key generation function selects a random integer $a_i$ as the signing key and computes $g^{a_i}$ as the verifying key for the consensus node $i$.

\textit{Signing}: Each consensus node $i$ computes its hashed identity and location information $m_i$ as $h_i = H (m_i)$, where $H$ is a hash function such as SHA-256 algorithm \cite{rachmawati2018comparative}. Then, this consensus node generates the one-time signature $\delta_i = {h_i}^{a_i}$ and sends it to the upper layer blockchain network.

\textit{Verifying}: Given the one-time signature $\delta_i$ and the verifying key $v_i$, the upper layer blockchain network can verify that $e(\delta_i, g) = e(h_i, v_i)$. This holds because $e({h_i}^{a_i}, g) = e(h_i, g^{a_i}) = e(h_i, g)^{a_i}$ due to the Bilinearity. Based on the threshold requirement, the validity of the lower layer blockchain network $r$ is verified only if at least $t$-out-of-$n$ consensus nodes' one-time signatures $r_i~(1 \leq i \leq n)$ are verified, we present $(r_1, r_2, ..., r_n) {\xrightarrow{(t,n)} r}$.

\begin{figure}[t]
\centering
\includegraphics[width=0.488\textwidth]{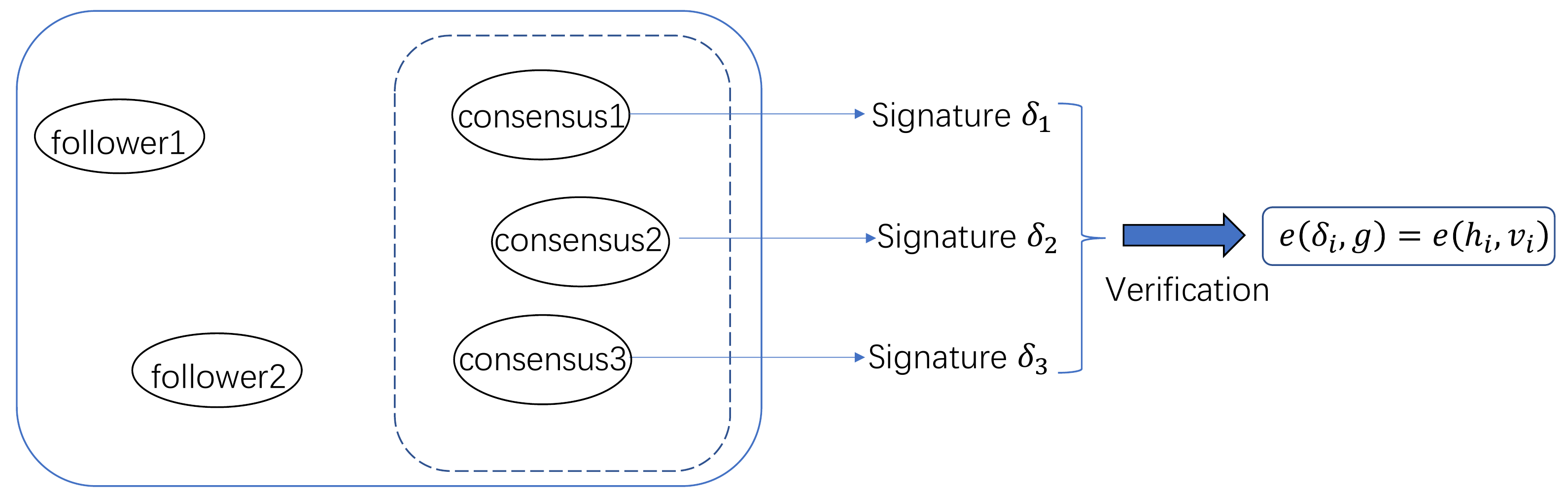}
\caption{T-of-n Threshold Signature Scheme.}
\label{fig:thresholdsig}
\end{figure}

As shown in Fig.~\ref{fig:thresholdsig},  $\it (t, n)$ threshold signature scheme has been applied based on the BLS signature scheme, where $1\leq t \leq n$.
For instance, If consensus nodes generated three different signatures related to the identity and location information. The upper layer verifier node, which verifies the generated signature $\delta_i$ for consensus node $\it i$, will follow the threshold of {\em 2 out of 3} to authenticate the identity and location information.  Also, the verifier node can apply the threshold signature scheme dynamically, such as the {\em 1 out of 3} rule, to provide more flexibility on the trust between the upper and lower layer blockchain network. 

Compared to the original {\it Raft} protocol, our proposed LH-Raft algorithm can tolerate the crash and byzantine fault. Classical {\it Raft} protocol requires that all the participant nodes are honest and conduct truthful action. By applying the {\it t-of-n} threshold scheme, the verifier node can check and verify the generated signatures independently with fault-tolerance property and protect the privacy of consensus nodes.





\subsection{LH-Raft Consensus Construction}
We describe the process of the location-based hierarchical raft protocol in Algorithm \ref{algorithm1}. The algorithm contains six primary phases: lines 2-7 shows the startup of election by a {\it follower} node; lines 8-11 describes the nearby node $n_i$ ($\forall$ $n_i$; $n_i$ $\in~N$) sends its~$CGF_i$ score to the {\it follower}  node; lines 12-15 presents the {\it follower}  node $F$ sorts the $CGF_i= \mathcal{\tau}_{i} + \mathcal{\sigma}_i$ score~$\forall i$ $~\in~\mathcal{N}$, and chooses the top M scores to form the candidate group $C$ and broadcasts the candidate group $C$ information, lines 16-18 indicates the confirmation of candidate group, lines 19-22 shows the confirmation of the elected {\it leader}, and finally lines 23-26 presents the current {\it leader}  switches its role to the {\it follower}.

\begin{algorithm}[ht]
\begin{algorithmic}[1]
\State {\bf OUTPUT:}  The $Leader$ of LH-Raft Protocol  
\State {\textbf{START UP}} {\em Follower node $F$ begins the Election.} 
\State {$F$ sends $FormGroup$ request to all nearby nodes $N$;}
 \State $F$ waits for messages in a time period T;
 \State {$F$ gets $CGF$ scores from all nearby nodes $N$;}
  

  \If {no $answer$ from $N$ within time $T$} 
  \State  $F$ restarts  $Election$ procedure;
  \State {\textbf{END START UP}} 
 \State {\textbf{UPON EVENT}} {\em Nearby nodes $N$ receive the FormGroup message:}
 \State  $N$ calculate  $CGF_i = (\mathcal{\tau}_{i} + \mathcal{\sigma}_i)$ for each $n_i$;
 \State each $n_i$ sends $CGF_i$ score to $F$;
 \State {\textbf{END UPON EVENT}}
 \State {\textbf{UPON EVENT}} {\em Follower node $F$ receives the $CGF$ score from all nearby nodes $N$:}
 \State $F$ sorts all $CGF_i$ scores 
 and chooses top $M$ nodes;
 \State $F$ broadcasts candidates group $C$ to all followers $F$;
  
\State {\textbf{END UPON EVENT}}
\State {\textbf{UPON EVENT}} {\em Nearby nodes $N$ receive the Follower message:}
 \State $N$ accept the nodes $M$ in candidate group $C$;
   \State {\textbf{END UPON EVENT}}
\State {\textbf{UPON EVENT}} {\em Candidates $C$ gets timeout and starts $Election$:}

   \If{$C_i$ receives $majority$ votes}
  \State $C_i$ becomes $Leader$ and broadcasts $Leader$ confirmation message to $Followers$ and $Candidates$;
  \Else
  \State $C_i$ waits for $Leader$ message from other $C$;
     \State {\textbf{END UPON EVENT}}
    \State {\textbf{UPON EVENT}} {\em Leader $L_i$ discovers nodes with higher term:}
    \State $L_i$ accepts other new node $L_j$ as the $Leader$;
    \State $L_i$ switches its role to $Follower$;
 \State {\textbf{END UPON EVENT}}
\caption{Location-based Hierarchical Raft Protocol}
\label{algorithm1}
\end{algorithmic}
\end{algorithm}



{Each IoT device must periodically submit its location, reputation, and timestamp information in the LH-Raft protocol. We utilize the Crypto-Spatial Coordinates (CSC) mechanism to connect geographical information of IoT devices~\cite{lao2020g}. One CSC consists of the hashed geographic location information and the smart contract header address. The resolution of CSC is about 1-square-meter around the specific area. CSC mechanism helps the IoT device to build immutable property to its physical location.  With CSC, IoT devices can have access to their historical location information. After the qualified IoT node is elected as the {\it leader} node, it will start to validate and generate a new block and manage blockchain new transactions based on the LH-Raft consensus protocol. If there is a missing block caused by the {\it leader} node, the current {\it leader} node will be removed from its {\it leader} status.}



To become the qualified {\it candidate} nodes, the {\it follower} nodes need to satisfy the geographic location requirements. Therefore, the LH-Raft protocol will check the geographic information of {\it follower} nodes periodically. It will determine if the {\it follower} nodes are within a particular geographic area and whether the node changes its location over some time. If a node's geographic location information has been changed significantly over the past period {\it t}, it will be removed from the {\it candidate} group. 

{To guarantee the system safety, we require that only one {\it candidate} node joins or leaves at the same time, which is similar to the classical Raft protocol. A majority of {\it candidate} nodes will reach the consensus on {\it candidate} group changes.  For each operation, local consensus on new candidate group formation must occur, and a majority of {\it candidate} nodes will know about the candidate group size changes. As the LH-Raft leader election process only differs in the $CGF$ and threshold signature generation phase from classical Raft protocol when a {\it leader} is selected, safety is preserved for {\it candidate} nodes join or leave, just as proven by Ongaro~\cite{ongaro2014search}.} 

{Even if the {\it candidate} nodes leave the network silently, the candidate group sizes may be decreased. As a result, consensus procedure or leader election results may be based on the candidate group sizes, which are smaller than necessary. However, even using smaller candidate group sizes will not lead to two {\it leader} nodes being elected and hinder safety property.}


{We also require the election restriction, which guarantees that a {\it candidate} node never wins the {\it leader election} if it does not have all committed entries in its log. In each sub-layer network, at least one node ({\it leader}) will have the latest committed entry. If a {\it follower} node receives a {\it RequestVotes} from a {\it candidate} node which is behind in the log (a smaller term number, or same term number but smaller index), it will not grant its vote.}
 
{To guarantee the system liveness, no concurrent transaction exists in our protocol. Otherwise, the newly generated transaction can be overwritten or cause the fork. If most {\it candidate}  nodes leave the system silently, then there should be an active {\it leader} to detect and report the status change. If the {\it leader} fails before committing the new transaction log or the nodes that silently left the system, the remaining {\it candidate} nodes can not elect a new {\it leader}. To address the above problem, our proposed scheme adopts the  $heartbeat$ message between different participant nodes as a failure detector for the system, similar to the original Raft consensus protocol.}

{As the {\it global consensus} is based on the {\it local consensus}, similar system liveness conditions will be applied. If the conditions of liveness in LH-Raft do not hold within a {\it candidate} group, we consider the {\it candidate} group has failed. For instance, if a majority of {candidate} nodes have failed, then the {\it local leader} will not be elected and cannot append global log entries and block the consensus process for the upper-layer network. To guarantee the liveness at the global level, liveness must first be guaranteed for both {\it intra-layer} consensus in enough {\it candidate} nodes for the {\it intra-layer} consensus to continue.}

\subsection{Local and Global Log Replication with Network Consistency}


The goal of LH-Raft is to boost the throughput and reduce the execution time of the consensus process in  large distributed systems. LH-Raft consists of two levels of log replication: {\it local log replication} with intra-layer {\it candidate}
 nodes, which have lower network delay and latency, and {\it global log replication} on batches of locally committed transactions to keep system consistency.

\begin{algorithm}[t]
\label{alg:local-leader}
\SetAlgoLined
\LinesNumbered
\SetKwInOut{Input}{Input}
\SetKwInOut{Output}{Output}
\Input{New localIndex $k$ has been received for the global log}
\Output{Update globalIndex $i$ for new entries $e$}
\While {there exists a new entry $k = localIndex +1$ has been received for the global log}
{Execute intra-layer consensus for global log replication;}
\eIf{$e.newInsert = leader$}{insert it to log \;}{wait for other leader \;}
The system updates globalIndex $i$ for new entries $e$ \;
\caption{Local Leader Log Replication}
\end{algorithm}

\vspace{3mm}
\noindent

In addition to the global blockchain ledger, each {\it leader} in the sub-layer blockchain replicates the local  transaction log. The local blockchain transaction log serves two purposes: first, buffering new entries for the global transaction log. Second, state replication for the inter-layer. Within each intra-layer, IoT devices propose new event entries be first placed in the local blockchain log. Periodically, the {\it leader} of the intra-layer blockchain proposes a batch of local entries to be committed and saved to the global transaction log. Batches could be created and proposed based on how many new entries have been saved in the local blockchain log.


In inter-layer consensus, the {\it leader} of each sub-layer blockchain network is elected by the local candidates' group members. Next, all the local {\it leaders} from the inter-layer establish the intra-layer log replications. All local {\it leaders} elect a global {\it leader} based on our proposed threshold signature schemes. Once the local {\it leader} is elected, it submits a specific log entry for the local blockchain log. The purpose of this log entry is to replicate the local {\it leader's} status in the inter-layer consensus process. Once the new entries are committed to the local blockchain log, the local {\it leader} will insert them into a global log for future replication.

{As shown in Algorithm 5, the global log replication is run through the intra-layer consensus process when the local {\it leader} receives a new proposal for the new global entry with $AppendEntries$ message. Local {\it leaders} who contain the $localIndex$ indicating which entries are committed in the global blockchain log. Local {\it leaders} include their $localIndex$ in the $AppendEntries$ messages to let all {\it follower} nodes at the sub-layer blockchain know which new entries have been committed in the global blockchain transaction.}

\begin{algorithm}[t]
\label{alg:merge-network}
\SetAlgoLined
\LinesNumbered
\SetKwInOut{Input}{Input}
\SetKwInOut{Output}{Output}
\Input{Sub-layer network $SN_1$, sub-layer network $SN_2$, integer $n$ and $m$}
\Output{Merged upper layer network $MN$}
\While {the upper layer leader has been elected by all sub-layer local leaders and other candidate nodes}
{Build merged upper layer network $MN$ from $SN_1$ and $SN_2$;}
\For{$int$ $i = 0;$  $i < n;$  $i++$}{merged[i] = N[i];}
\For{$int$ $i = 0;$ $i < m;$ $i++$}{merged[n + i] = M[i];}
\Comment{Replicate nodes of $SN_1$ and $SN_2$ one by one
     to upper network merged[].}
     
buildUpperLayerNetwork $(merged$ $MN$, $n + m);$


\caption{{Merge Sub-layer Network}}
\end{algorithm}

\vspace{3mm}
\noindent

{In LH-Raft consensus protocol, after all {\it local leaders} and other {\it candidate nodes} elect the upper layer {\it leader}, it can merge and construct all leaf layer networks into upper-layer network structure, and the merge operation only happened once between the middle and top layer blockchain network. Ideally, we consider all sub-layer networks as the heap structure, a specialized tree-based data structure.  We adopted the max heap property, and the $CGF$ score of {\it leader} node is always greater than or equal to those of the {\it follower} and {\it candidate} nodes.}

{As shown in Algorithm 6, the function of merging sub-layer network is executed through the inter-layer consensus process when the upper {\it leader} has been elected by all sub-layer {\it local leaders} and other {\it candidate nodes}. The algorithm builds the merged upper layer network $MN$ from sub-layer networks $SN_1$ and $SN_2$ (In a real-world scenario, it can have multiple sub-layer networks). Line 4-9 presents the replicating nodes of sub-layer $SN_1$ and $SN_2$ one by one to the upper-layer network. Line 11 builds the upper-layer network from the sub-layer networks $SN_1$ and $SN_2$.}

\section{Theoretical Analysis}
This section analyzes the LH-Raft protocol from three perspectives: performance, overhead, and fault-tolerance.

\subsection{Performance Analysis}

One of the main innovations of the LH-Raft protocol is to pre-elect the {\it candidate nodes} to form the consensus group to elect the {\it leader} to run the LH-Raft protocol, which can dynamically adapt to the geographic change of IoT devices. The system performance improves significantly because the consensus group's size in LH-Raft is markedly smaller than in IoT networks.



Let {\it n} and {\it c} represent the number of total IoT devices and {\it candidate nodes} in LH-Raft, respectively.  {\it p} is the processing power of the IoT node, which indicates that the node can receive and process {\it p} messages every second~\cite{lao2020g}. Compared to the classical Raft protocol, a node needs to receive the majority approved message to become the {\it leader node} (for instance, we pick $(2*n) /3$ messages here to reflect the threshold in our proposed threshold signature scheme). Consequently, it takes at least $(2*n) /(3*p)$ seconds to complete the leader election process. The consensus procedure in classical Raft is in the order of $n/p$, whereas the consensus process in LH-Raft is in the order of  $c/p$. As a result, the time to accomplish the consensus could be reduced to $c/n$. Therefore, the larger the total number of IoT devices to geographic-based candidates, the greater the overall system performance enrichment it can achieve.

\subsection{Overhead Analysis}

One of the drawbacks of the classical Raft's insufficient scalability is the communication cost problem. A node needs to broadcast the {\it RequestVotes} message to all the participating nodes and then get the {\it Reply} message from the majority of the IoT nodes. Therefore, the communication overhead of Raft is ${\it O(n^2)}$, where {\it n} represents the total number of IoT devices. 

Our proposed LH-Raft protocol could reduce the size of {\it candidate} group and the communication cost. Since the node in LH-Raft only sends the message to other physically nearby {\it candidate} nodes, the communication overhead of LH-Raft is ${\it O(c^2)}$, where {\it c} stands for the {\it candidate} group. As a result, the LH-Raft algorithm can reduce the communication cost in the order of $(c^2)/ (n^2)$. The larger the number of IoT nodes to {\it candidate} nodes is, the larger the decrease of  the communication overhead it can achieve.

In classical Raft protocol, the communication overhead is $O(n^2)$. In contrast, the LH-Raft protocol will select the {\it candidate} nodes based on their reputation score and the geographic information, and only the limited number of {\it candidate} can participate in the election process to elect the {\it leader} node. Consequently, the overall communication overhead is $O(c^2)$, where $c << n$. 


\subsection{Fault-tolerance Analysis}

Traditional IoT-blockchain system usually suffers from the bad activities launched by malicious nodes. In a permissionless blockchain network, a malicious node can spawn massive dishonest nodes. For instance, if enough dishonest nodes control the consensus group (1/3 in PBFT, 1/2 in PoW), malicious nodes can tamper and fork the ledger's transaction information.

{The LH-Raft consensus protocol could tolerate $f$ failures with $2f + 1$ total nodes under non-byzantine conditions such as node crashes, network delays, packet omission, and record tampering issues. For the byzantine fault, our proposed scheme could tolerate $f$ failures with $3f+1$ nodes, and the $t-of-n$ threshold signature scheme could also help with the malicious attack.}


{Our LH-Raft mechanism requires the {\it follower} nodes to upload their geographic information periodically to address the above issue. We assume that different IoT devices cannot declare the exact geographic location with the same timestamp, limiting the total number of IoT nodes participating in the malicious attacks. 
In addition, LH-Raft forms a nodes group from IoT devices located in a relatively small geographic region; local nodes may verify each others' locations and report fake geographic areas submitted by malicious nodes.}


{For instance, any geographic data submitted from this area will be considered fake  if there is no IoT node in a specific geographic location. Next, the threshold signature scheme can reach a consensus if consensus nodes' signatures have satisfied the verification process and meet the threshold requirement. As a result, malicious nodes cannot generate valid signatures and pass the verification process. All malicious nodes can generate fake geographic information but cannot tamper with or forge messages sent by other honest nodes without valid signatures. Finally, we convert the voting process in classical Raft protocol to the signature verification with a dynamic threshold scheme to help with the byzantine fault tolerance.}

\section{Experiments and Evaluations}
We developed the IoT blockchain system prototype with the LH-Raft consensus protocol and compared the LH-Raft protocol with the classical Raft protocol. 

\begin{table}[t]
  \begin{center}
    \caption{Parameter Setting}
    \label{tab:table2}
    \begin{tabular}{l|r}
      Parameter Name & Value (Range) \\ 
      \hline
      Classical Raft candidate nodes $cr$ & 20-400\\
      LH-Raft candidate nodes $c$ & 20-2000 \\
      Total participating nodes $n$ & 1-5000 \\
      Decision variable $X_{Vi}$ & 0-1 \\
      Identity information $m_i$ & 0-256bit \\
      Threshold parameter $t$ & 1-n \\
      Total number of verified nodes $k$ & 1-t \\
    \end{tabular}
  \end{center}
\end{table}

\subsection{Experimental Setup}
The initial blockchain network contains four consensus nodes and then extends to a larger number of nodes to compare LH-Raft latency and communication overhead with the classical Raft protocol. The threshold signature scheme is developed and tested on the Hyperledger Ursa, a cryptographic library for blockchain applications. The classical Raft protocol has {\it cr} candidate nodes to participate in the leader election process. In contrast, the LH-Raft protocol has {\it c} candidate nodes participating in this process. In both experiments for the consensus latency and communication cost, we set {\it cr = 2c} as the pre-condition. Note that in large-scale IoT networks {\it cr} can be orders of magnitudes larger than {\it c}, which significantly hinders the classical Raft performance against LH-Raft. 

In this section, the parameters used in the experiments and evaluations are shown in Table \ref{tab:table2}.

\subsection{Candidate Group Formation Cost}
In this subsection, we evaluate the impact of changes in candidate group size and the total number of participating nodes on the execution time of the candidate group formation mechanism. IP-CGF is optimally solved utilizing Python 3.0 mathematical libraries. 

\begin{figure}[t]
\centering
\includegraphics[width=0.38\textwidth]{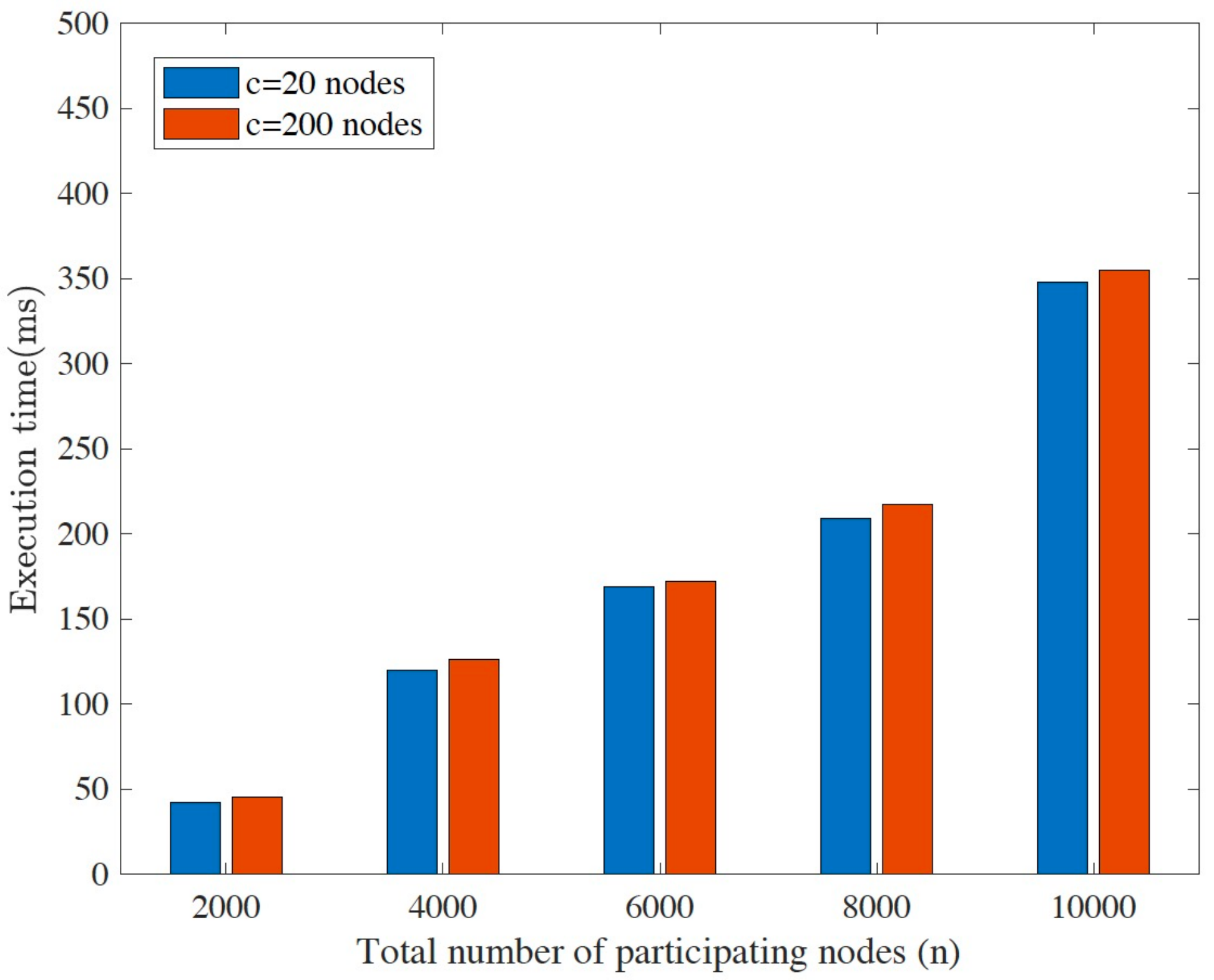}
\caption{Candidate group formation cost vs. the total number of participating nodes.}
\label{fig:cgfscore1}
\end{figure}

As shown in Fig.~\ref{fig:cgfscore1}, we changed the total number of participating nodes from 2000 to 10000, and the size of the candidate's group $(c)$ is configured as 20 and 200 nodes. The execution costs for 2000, 4000, 8000, and 10000 participating nodes are 42ms and 45ms, 120ms and 126ms, 169ms and 172ms, 209ms and 217ms, and 348ms and 355ms for $c=20$ and $c= 200$ nodes, respectively. When the total number of participating nodes increases, the execution time to form the candidate's group also increases significantly. Note that the execution time result is close when $c$ is 20 and 200 nodes in each experiment round. 
It shows that the size of the candidate group has relatively less impact on the execution time than the total number of participating nodes when $c$ is relatively small. The reason is that the time complexity of the IP-CGF solution algorithm is $O(n~log(n))$, where $n$ is the total number of participating nodes.


\begin{figure}[t]
\centering
\includegraphics[width=0.38\textwidth]{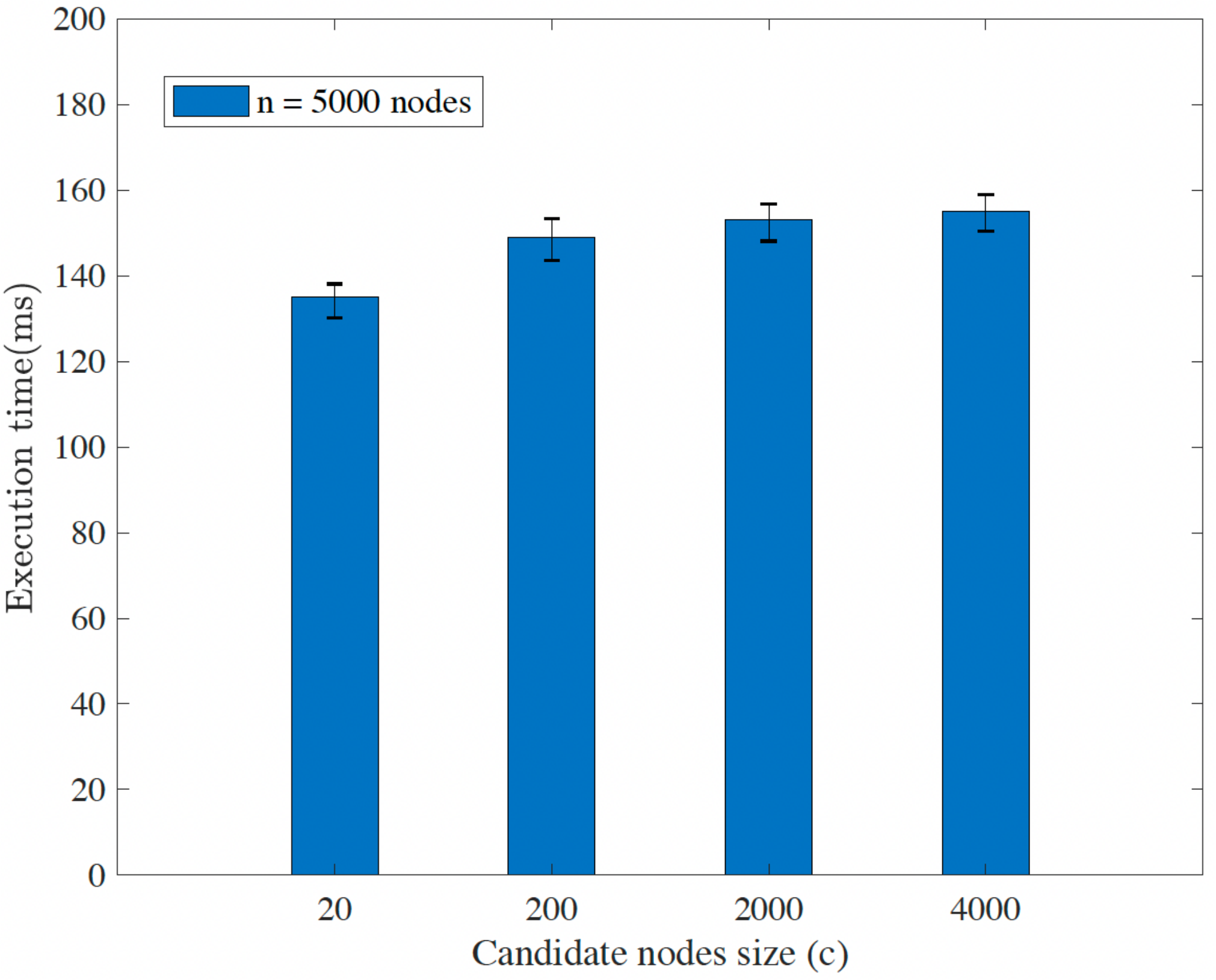}
\caption{Candidate group formation cost vs. candidate group size when the total number of participating nodes is 5000 constantly.}
\label{fig:cgfscore2}
\end{figure}


In Fig.~\ref{fig:cgfscore2}, we modify the size of candidates group number $c$ from 20 to 4000, and the total number of participating nodes $(n)$ is configured as 5000 nodes. 
As shown in Fig.~\ref{fig:cgfscore2}, the average IP-CGF execution costs in 20, 200, 2000, and 4000 candidate group nodes are 135ms, 149ms, 155ms, and 159ms, respectively. Blue color represents different execution costs under different candidate nodes' sizes, and the black error bar indicates the confidence interval. The execution time increases slightly when the candidate group size increases from 200 to 4000 nodes. The close execution time when the threshold $c$ is 20, 200, 2000, and 4000 nodes show that the candidate group's size has a non-significant impact on the execution time of the proposed candidate formation mechanism by IP-CGF.



Comparing the results in Fig.~\ref{fig:cgfscore1} and Fig.~\ref{fig:cgfscore2}, the execution time is higher when increasing the total number of participating nodes and slightly higher with the increased size of the candidate group with fixed participating nodes. We argue that 20 and 200 can be feasible for the candidate group in a real-world application with our proposed hierarchical blockchain architecture with the most IoT-blockchain applications.

\begin{figure}[t]
\centering
\includegraphics[width=0.39\textwidth]{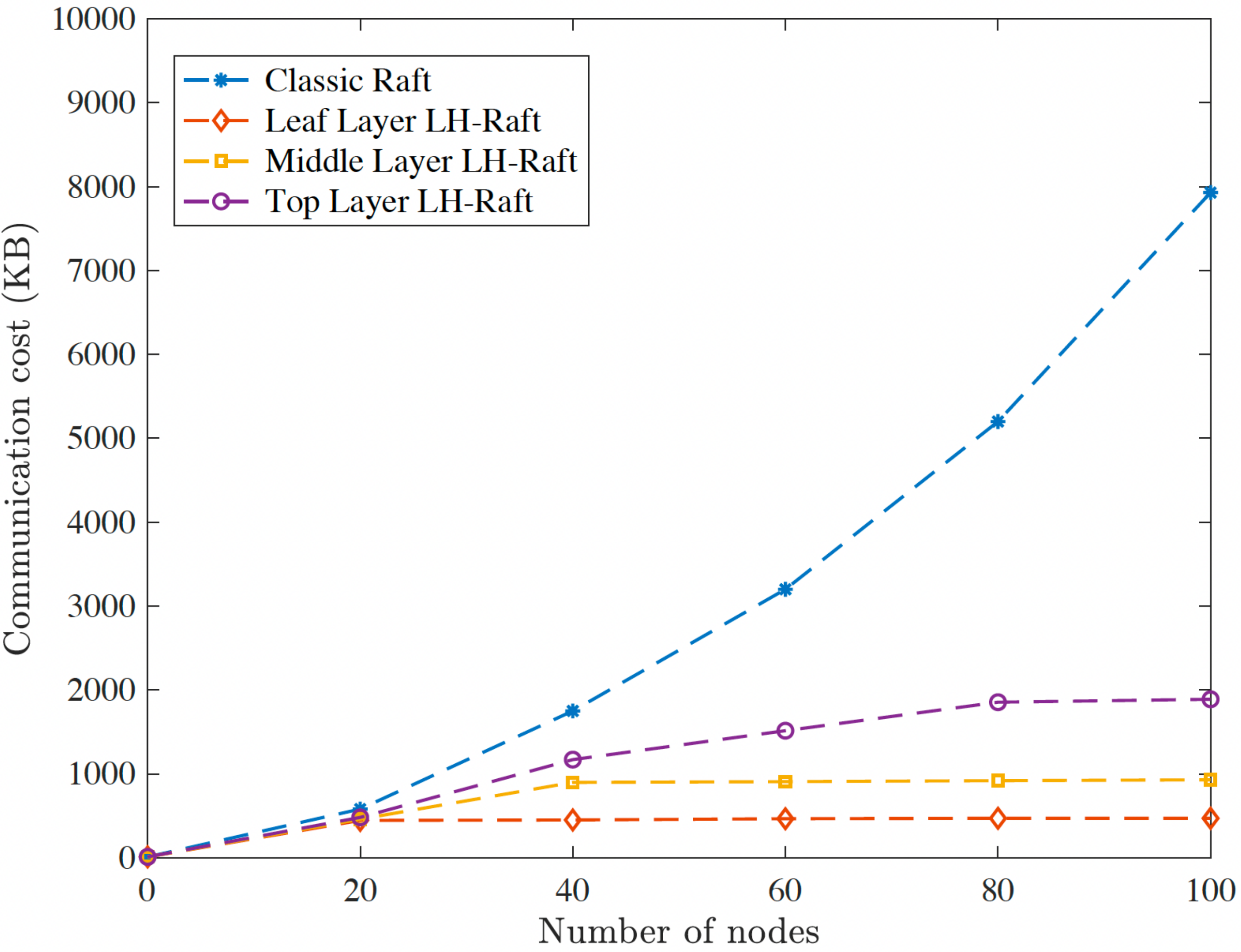}
\caption{{Comparison between LH-Raft sub-layer and classical Raft communication costs for each transaction.}}
\label{fig:commcost}
\end{figure}

\begin{figure}[t]
\centering
\includegraphics[width=0.39\textwidth]{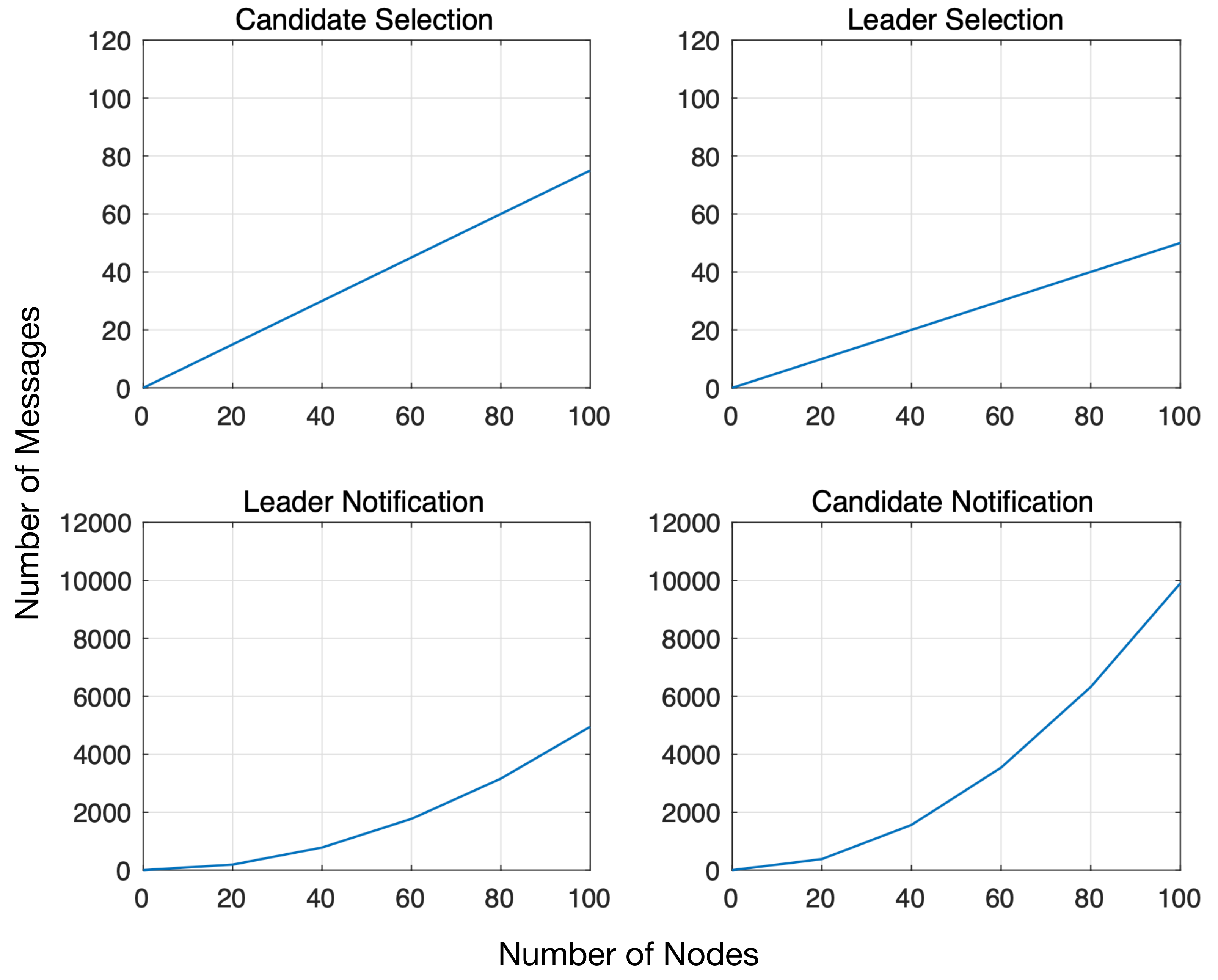}
\caption{{Number of messages vs. number of nodes from each step in LH-Raft consensus.}}
\label{fig:messagecomm}
\end{figure}

\begin{figure}[t]
\centering
\includegraphics[width=0.39\textwidth]{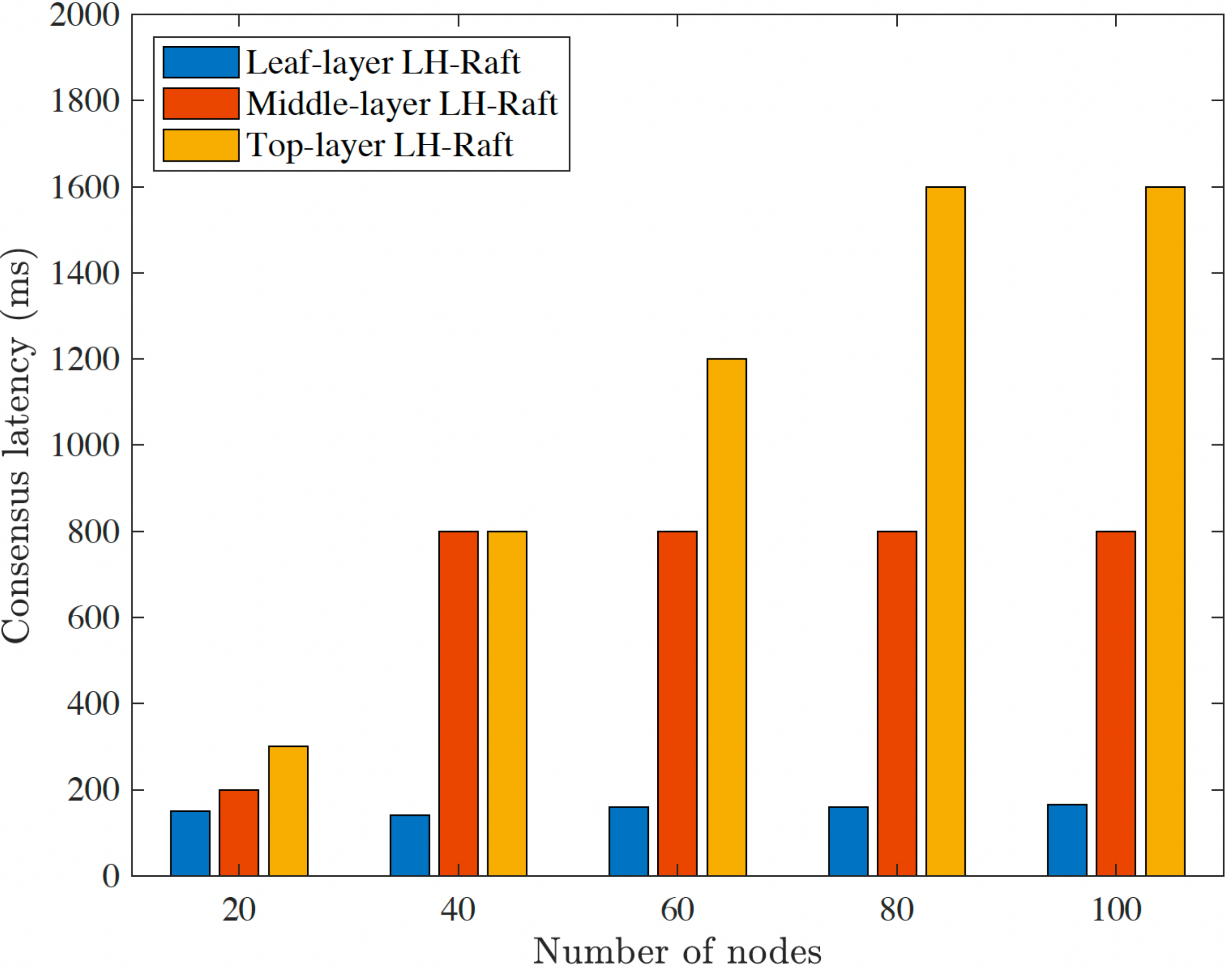}
\caption{{Sub-layer's consensus latency in LH-Raft.}}
\label{fig:raftlatency}
\end{figure}

\begin{figure}[t]
\centering
\includegraphics[width=0.39\textwidth]{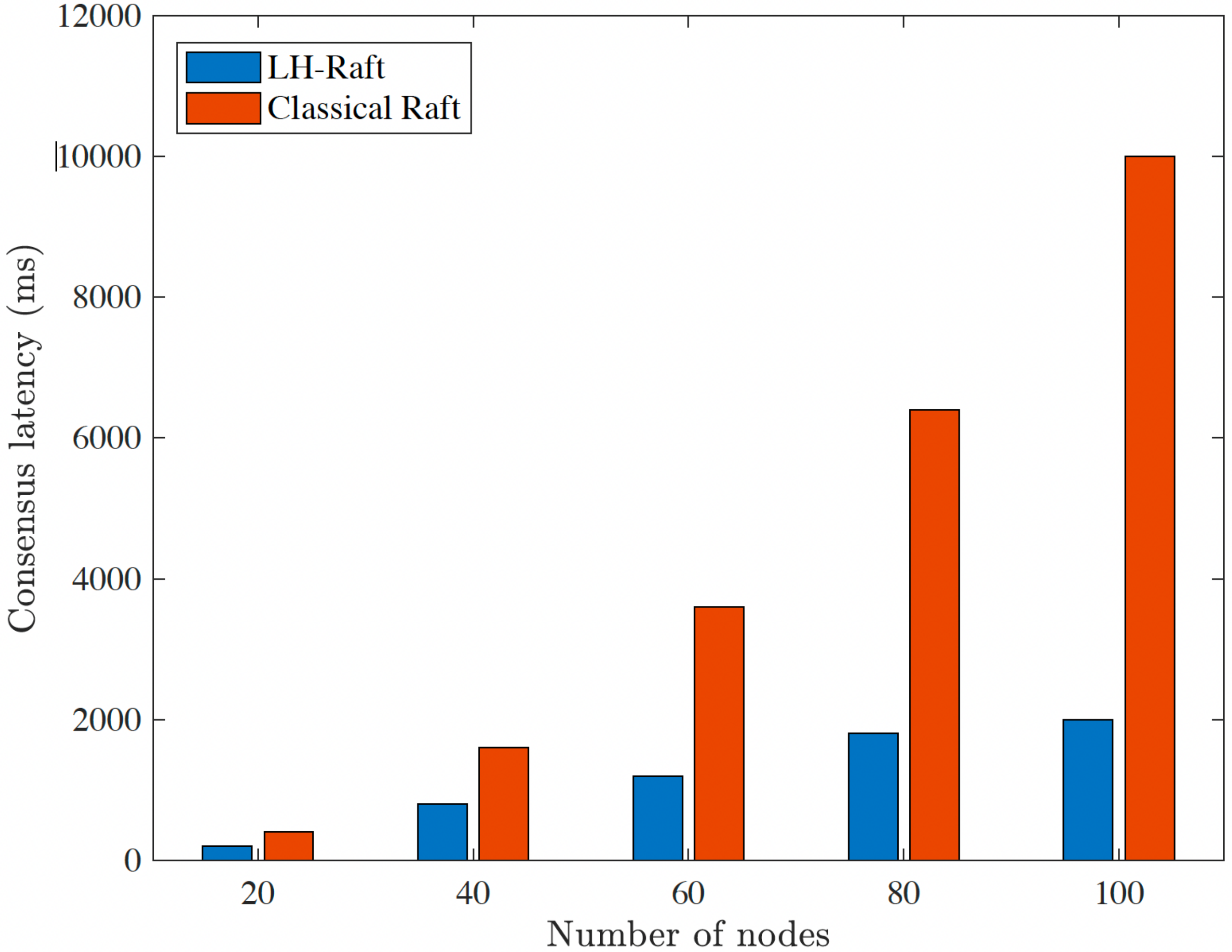}
\caption{{Comparison between LH-Raft and classical Raft consensus latencies.}}
\label{fig:raftoverlatency}
\end{figure}

\subsection{Communication Cost}

{Our proposed LH-Raft protocol can reduce the communication cost significantly in multiple network layers when the number of IoT devices is large. We emulate the leaf-layer, middle-layer, and top layer blockchain's {\it candidate} nodes to be 20, 40, and 80 at maximum and evaluate the communication cost for all nodes in one single blockchain transaction. It elects the {\it leader} from the {\it candidate} nodes and sends the transaction. As shown in Fig.~\ref{fig:commcost}, communication cost in the classical Raft consensus protocol keeps growing when the number of nodes increases. Moreover, the larger the number of nodes participating in the system, the more significant the increase in communication cost. However, for the LH-Raft protocol, the communication cost grows linearly from 0 to 20 for the number of nodes in the leaf-layer blockchain network and reaches the upper bound of about 470kb since the leaf-layer blockchain has the maximum capability for {\it candidate} nodes. Similarly, the communication cost also grows linearly from 0 to 80 for the {\it candidate} nodes in the top-layer blockchain network, which is bounded at 1850kb when the number of {\it candidate} nodes is beyond 80.}


{
The blue dashed line representing the classical Raft protocol has the 8000kb communication cost when nodes are 100. In contrast, our proposed LH-Raft consensus protocol can reduce the communication cost to 5.07$\%$ in the leaf-layer network and 22.16$\%$ even in the top-layer blockchain network, as observed from Fig.~\ref{fig:commcost}.}

\subsection{Message Passing}


By changing the number of nodes that participated in each state of the LH-Raft consensus protocol, Figure \ref{fig:messagecomm} indicates the total number of messages from each phase during LH-Raft consensus processing.  During the candidate selection phase ({\it follower} nodes form the candidate group), the number of messages increases slower than the leader selection phase ({\it candidate} nodes elect the {\it leader}) since our proposed LH-Raft algorithm selects the qualified {\it candidate} nodes based on their geographic and reputation score, which limits the candidate group's size.


Next, the leader notification phase ({\it leader} communicates with {\it follower} nodes) exchanges fewer messages than the candidate notification phase ({\it candidate} nodes communicate with {\it follower} nodes) because the {\it leader} node conducts 1-to-$n$ communications to all {\it follower} nodes. The candidate notification messages will increase faster due to the $c$-to-$n$ communications for all {\it candidate} nodes communicate with {\it follower} nodes.



{Compared to the original Raft consensus protocol, our proposed LH-Raft algorithm reduces the size of the candidates' group and partitions the blockchain network into hierarchical architecture.  As a result, the total number of message communication will decrease due to the smaller candidates group and multiple local {\it leader} nodes, and the system has one global {\it leader} node in the top-layer network. Moreover, our system could tolerate the faulty messages by incorporating the threshold signature scheme with local and global log replication schemes.}
\subsection{Consensus Latency}

{This subsection compares the consensus latency results of LH-Raft with the classical Raft protocol. The consensus latency is the time when executing the leader election process in the LH-Raft and classical Raft. We pick one {\it follower} node randomly to join the {\it candidate} group per sub-layer blockchain.  As shown in Fig.~\ref{fig:raftlatency} and Fig. 10, the LH-Raft consensus protocol in different layers showed significant latency improvements over the classical Raft by increasing the number of nodes (e.g., {\it 2.1x} consensus latency decrease for 20 leaf-layer networks). We measured the numerical consensus latency results of leaf-layer, middle-layer, and top-layer against the classical Raft protocol. Compared to the classical Raft, LH-Raft selects a smaller number of {\it candidate} nodes based on the geographic information belonging to each sub-layer network. When nodes increase from 20 to 100, consensus latency grows exponentially in the classical Raft protocol. The consensus latency in sub-layers of LH-Raft is bounded by the number of {\it candidate} nodes.}



{By contrast, the LH-Raft consensus protocol performs better in terms of the consensus latency result. All {\it candidate} nodes can join the consensus group when the number of nodes is smaller than the maximal threshold of candidate groups (i.e., 40 for the middle-layer).  Consequently, when the number of {\it candidate} nodes grows from 1 to 100, the consensus latency increases 72$\%$ slower compared to the consensus latency results in classical Raft protocol in Fig. 10. However, once the number of {\it candidate} nodes reaches the maximum threshold, no more new nodes can join the candidate group, and the consensus latency will not increase anymore.}

\subsection{Sensitivity Analysis}

In this subsection, we conduct a sensitivity analysis to evaluate the effects of changing the {\it c/n} ratio on the performance comparison between the proposed LH-Raft and classical Raft protocols. As stated in the previous section, LH-Raft significantly reduces the consensus latency and communication cost, especially when the ratio of {\it c/n} is greater.

\begin{figure}[t]
\centering
\includegraphics[width=0.39\textwidth]{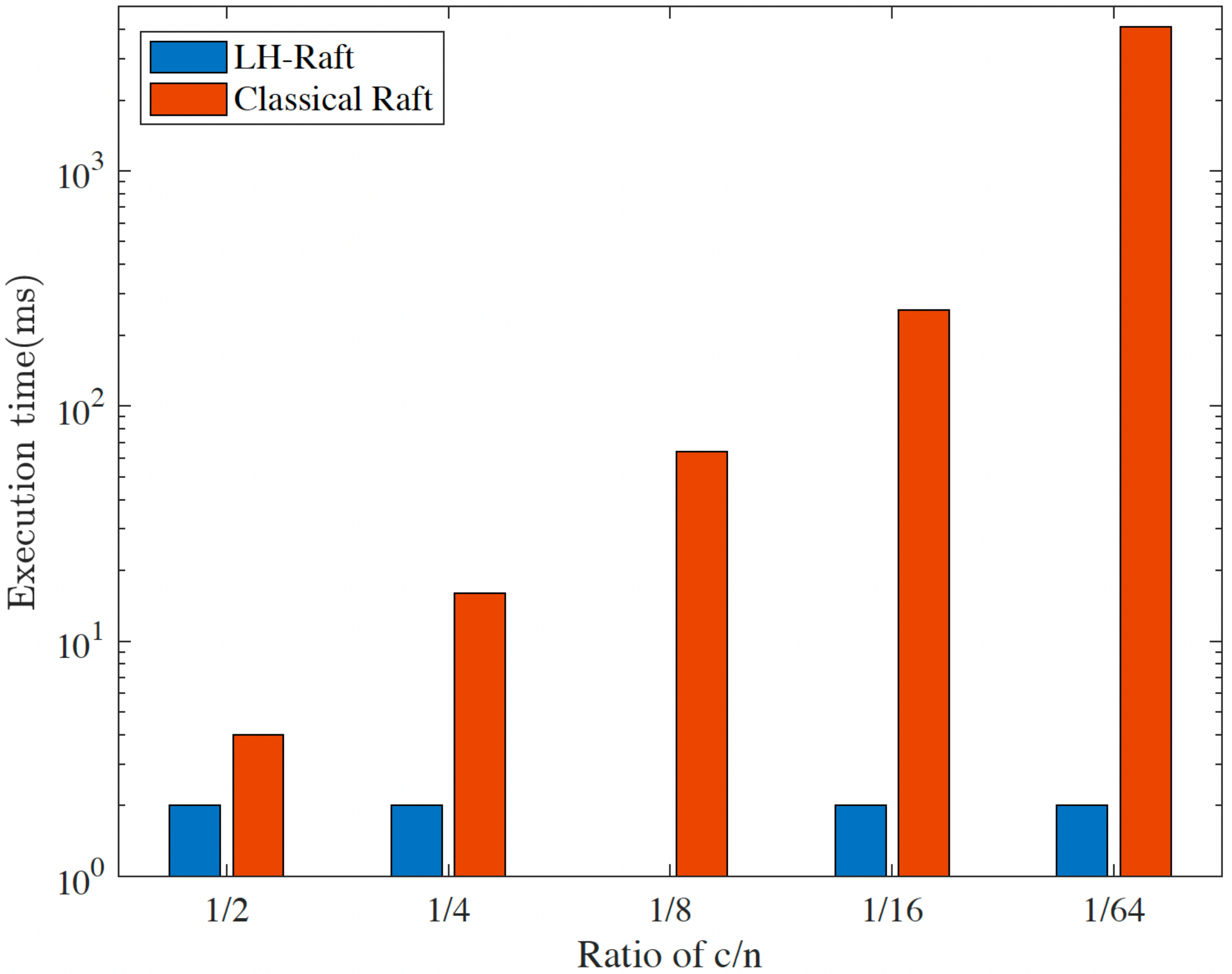}
\caption{Sensitivity analysis on the execution time of LH-Raft and classical Raft with different ratios of c/n.}
\label{fig:sensitivity}
\end{figure}

To evaluate the reduction of consensus latency, we showed multiple groups of experiments with different ratios of {\it candidate} nodes to total IoT nodes.  As shown in Fig.~\ref{fig:sensitivity}, our proposed LH-Raft consensus protocol showed significant performance improvements over the classical Raft by increasing the ratio of  {\it c/n}. For instance, when the ratio of {\it c/n} grows from {\it 1/2} to the {\it 1/64}, the consensus latency of the classical Raft protocol will grow exponentially in comparison with the proposed LH-Raft protocol.  
In {\it 1/64} case, the classical Raft protocol's consensus latency is beyond ${\it 10^3}$ while the LH-Raft protocol's performance remains at ${\it 10^0 to ~10^1}$ level.

\begin{figure}[t]
\centering
\includegraphics[width=0.39\textwidth]{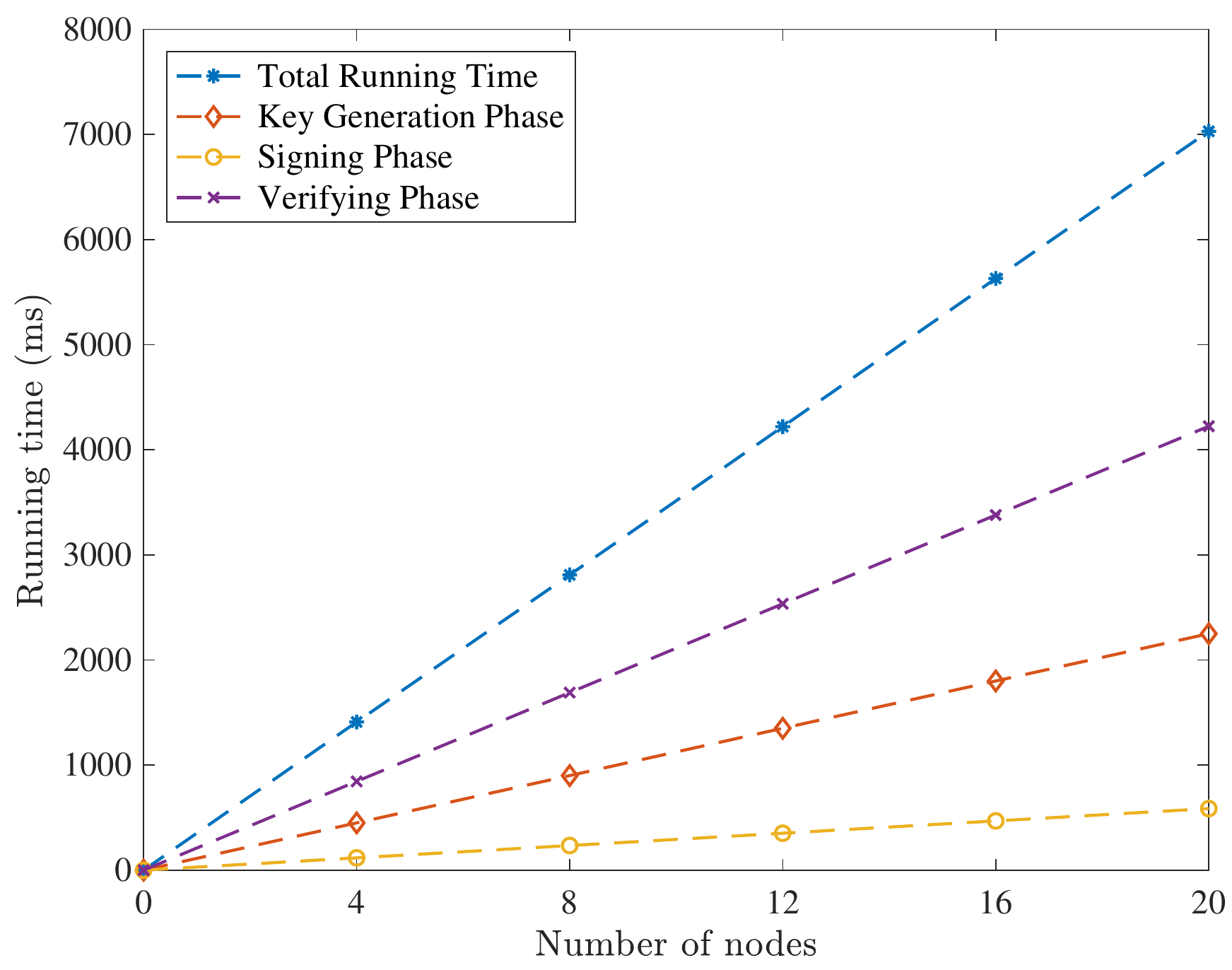}
\caption{Threshold signature scheme running time vs. number of consensus nodes.}
\label{fig:ursa_plot}
\end{figure}

\begin{figure}[t]
\centering
\includegraphics[width=0.484\textwidth]{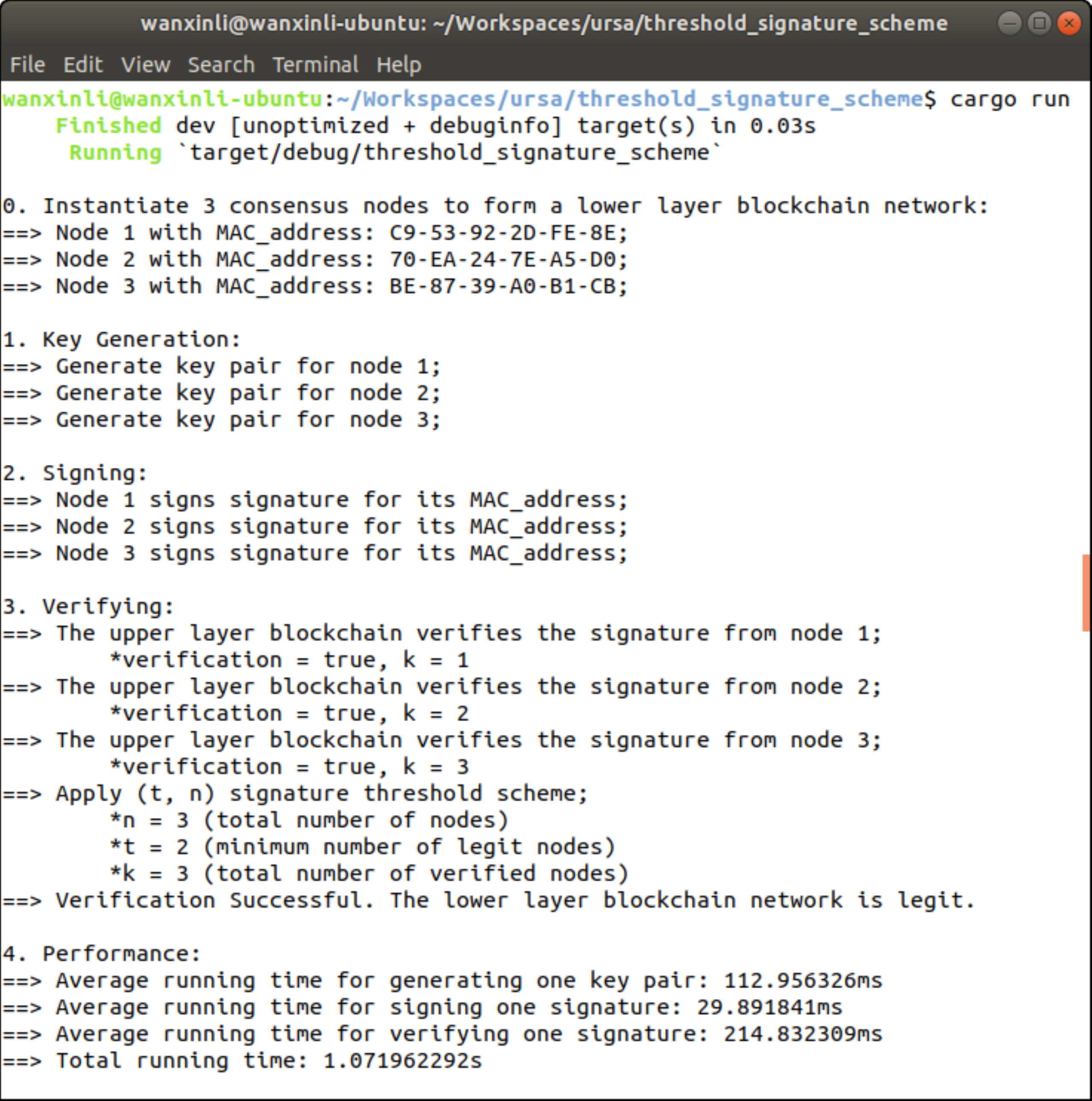}
\caption{Threshold signature scheme running procedure on the Hyperledger Ursa.}
\label{fig:ursa_window}
\end{figure}

\subsection{Threshold Signature Scheme}


We measure the performance of the threshold signature scheme by varying the number of consensus nodes from 4 to 8, 12, 16, and 20 in the lower layer blockchain network. As shown in Fig. \ref{fig:ursa_plot}, the total running time of each phase increases linearly with the increase in the number of consensus nodes in the lower layer blockchain network. Besides, the verifying phase takes additional time than the signing phase because the former requires the computation of pairings based on bilinearlity. In addition, our threshold signature scheme can offer constant running time for signing and verifying phases when varying the length of the identity information. The identity information $m_i$ from node $i$ is hashed to a fixed length of 256-bit value before signing the signature. Non-reliance on message length makes our scheme more flexible and efficient in verifying different types of identity information.


Our threshold signature scheme is developed on the Hyperledger Ursa. As shown in Fig. \ref{fig:ursa_window}, we first instantiated three consensus nodes to form a lower-layer blockchain network instance. Each node can be identified by its unique MAC address. Then, we follow Algorithms 1, 2, and 3 in constructing key generation, signing, and verifying phases. In this example, three signatures are signed by the consensus nodes from the lower layer blockchain network and confirmed by the upper layer blockchain network. For each signature, the average time for key pair generation, signing, and verifying are 113 ms, 30 ms, and 215 ms, respectively.

\section{Conclusion}
{This paper proposed a hierarchical and location-aware consensus protocol for IoT-Blockchain applications.
 The proposed LH-Raft mechanism forms localized consensus candidate groups based on their reputation score and the geographic information to elect the leaders. Our proposed LH-Raft is scalable by design and reaches consensus faster with lower network overhead and less communication cost than the original Raft protocol. 
 We conduct  numerical analyses based on the LH-Raft protocol  with the candidate group formation and message passing model. Also, we prototype the experiments by utilizing the Hyperledger Ursa cryptography library to evaluate the threshold signature scheme. The results indicate that the architecture is scalable 
 and suitable for large-scale and real-world IoT applications. 
 We plan to investigate the cross-chain consensus among heterogeneous blockchain systems for future work.
}
\ifCLASSOPTIONcompsoc
  \section*{Acknowledgments}
   This paper is a revised and extended version of ``A Location-based and Hierarchical Framework for Fast Consensus in Blockchain Networks" which is under review in 2021 4th IEEE International Conference on Hot Information-Centric Networking (HotICN 2021).  This work was partially supported by the Fundamental Research Funds for the Central Universities under the Grant G2021KY05101.
\else
  \section*{Acknowledgment}
   {This work was partially supported by the Fundamental Research Funds for the Central Universities under the Grant G2021KY05101, 2021-2024. This research is supported in part by the Guangdong Basic and Applied Basic Research Foundation under the Grant No. 2021A1515110286, 2021-2024, the Natural Science Foundation of Shaanxi Provincial Department of Education under the Grant No. 2022JQ-639, and a Federal Highway Administration grant: ``Artificial Intelligence Enhanced Integrated Transportation Management System", 2020-2023.} 
\fi


\ifCLASSOPTIONcaptionsoff
  \newpage
\fi

\bibliographystyle{IEEEtran}
\bibliography{sig.bib}


\section*{Biographies}
\vskip -2\baselineskip plus -1fil 
\begin{IEEEbiography}
[{\includegraphics[width=1.0in,height=1.25in,clip]{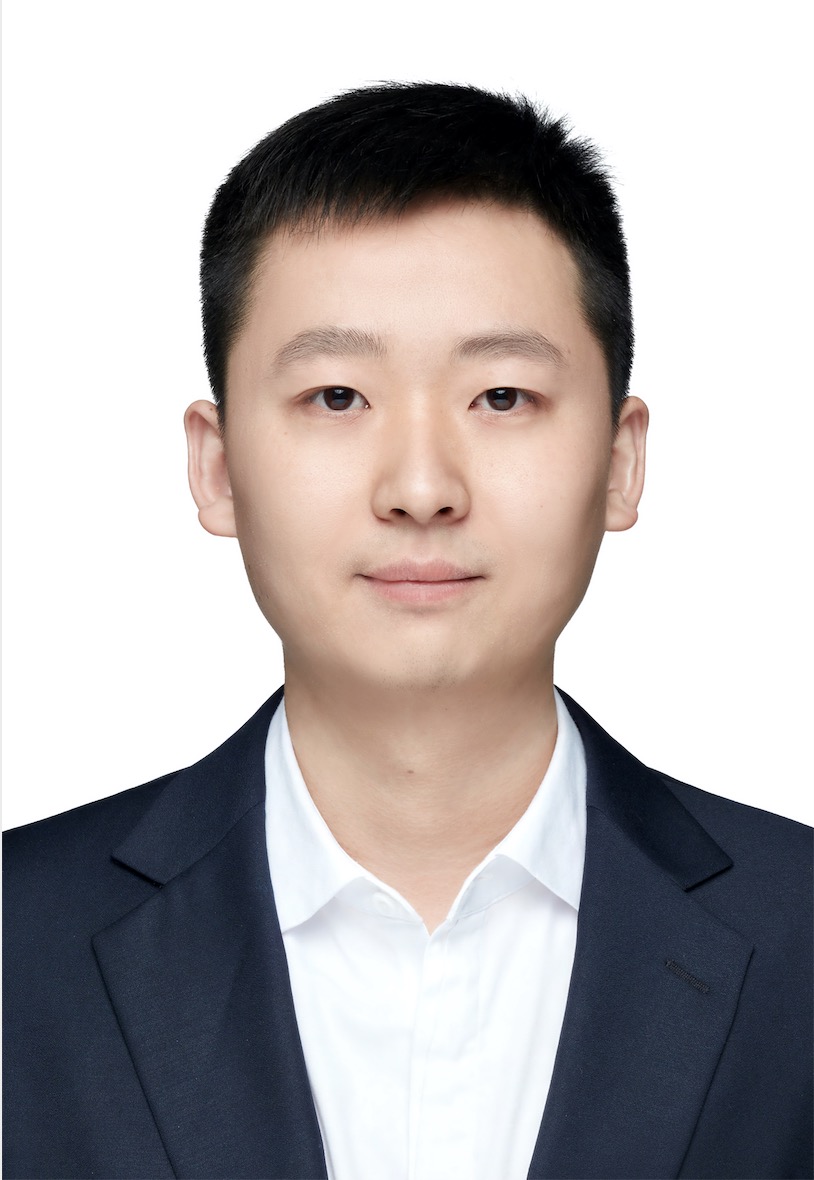}}]{Hao Guo} is currently an Assistant Professor with the School of Software at
the Northwestern Polytechnical University. He received the B.S. and M.S. degrees from the Northwest University, Xi'an, China in 2012, and the Illinois Institute of Technology, Chicago, United States in 2014, and his Ph.D. degree from the University of Delaware, Newark, United States in 2020, all in computer science.
His research interests include blockchain and distributed ledger technology, data privacy and security, cybersecurity, cryptography technology, and Internet of Things (IoT). He is a member of both ACM and IEEE.
\end{IEEEbiography}
\vskip -6pt plus -1fil
\begin{IEEEbiography}
[{\includegraphics[width=1in,height=1.25in,clip]{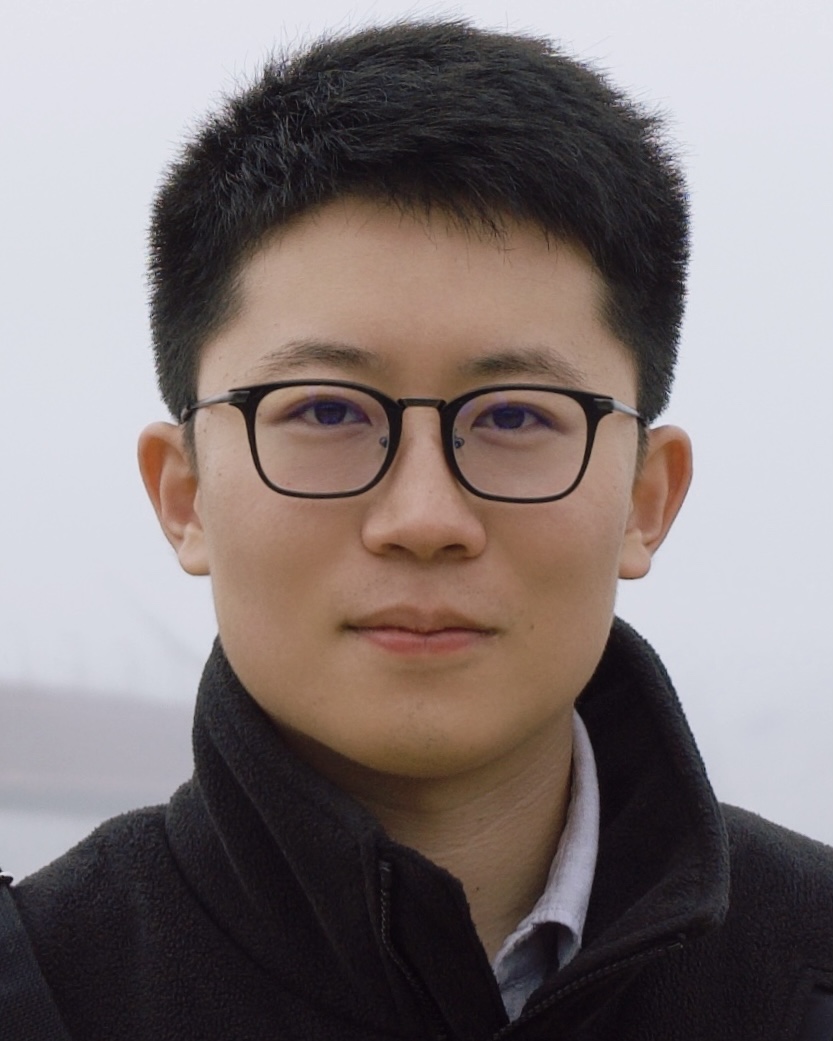}}]{Wanxin Li} (Member, IEEE) received the B.S. degree from the Chongqing University in 2015, and the M.S. and Ph.D. degrees from the University of Delaware in 2017 and 2022, respectively. He is a Lecturer with the Department of Communications and Networking, Xi'an Jiaotong-Liverpool University. His research interests are the security and privacy in blockchain, and blockchain-based architecture designs such as connected and autonomous vehicular networks, electronic health records and federated learning. 
\end{IEEEbiography}
\vskip -6pt plus -1fil
\begin{IEEEbiography}
[{\includegraphics[width=1in,height=1.25in,clip]{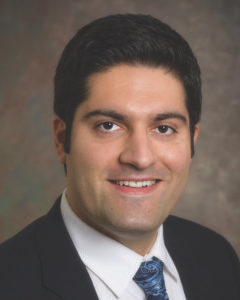}}]{Mark Nejad} is an Assistant Professor at the University of Delaware. His research interests include  network optimization, distributed systems, blockchain, game theory, and automated vehicles. He has published more than forty peer-reviewed papers and received several publication awards including the 2016 best doctoral dissertation award of the Institute of Industrial and Systems Engineers (IISE) and the 2019 CAVS best paper award IEEE VTS. His research is funded by the National Science Foundation and the Department of Transportation. He is a member of the IEEE and INFORMS.
\end{IEEEbiography}
\vskip -6pt plus -1fil







\end{document}